# PHYSICS-INTEGRATED MACHINE LEARNING: EMBEDDING A NEURAL NETWORK IN THE NAVIER-STOKES EQUATIONS. PART I

Arsen S. Iskhakov*, Nam T. Dinh**

*North Carolina State University, Department of Nuclear Engineering, Campus Box 7909, Raleigh, NC 27695 USA
aiskhak@ncsu.edu

**North Carolina State University, Department of Nuclear Engineering

## ABSTRACT

In this paper a physics-integrated (or PDE-integrated (partial differential equation)) machine learning (ML) framework is investigated. The Navier-Stokes equations are solved using the Tensorflow ML library for Python programming language via the Chorin's projection method. The methodology for the solution is provided, which is compared with a "classical" solution implemented in Fortran programming language. The Tensorflow solution is integrated with a deep feedforward neural network (DFNN). Such integration allows one to train a DFNN embedded in the Navier-Stokes equations without having the target (labeled training) data for the direct outputs from the DFNN; instead, the DFNN is trained on the field variables (quantities of interest), which are solutions for the Navier-Stokes equations (velocity and pressure fields).

To demonstrate performance of the framework, a case study is formulated: 2D lid-driven cavity with non-constant velocity-dependent dynamic viscosity is considered. A DFNN is trained to predict dynamic viscosity fields from velocity fields. The performance of the physics-integrated ML is compared with "classical" ML framework, when a DFNN is directly trained on the available data (fields of dynamic viscosity). Both frameworks showed similar accuracy; however, despite its complexity and computational cost, the physics-integrated ML offers principal advantages, namely: (i) the target outputs (labeled training data) for a DFNN might be unknown and can be recovered using the knowledge base (PDEs); (ii) it is not necessary to extract and preprocess information (training targets) from big data, instead it can be extracted by PDEs; (iii) there is no need to employ a physics- or scale-separation assumptions to build a closure model for PDEs. The advantage (i) is demonstrated in this paper, while the advantages (ii) and (iii) are the subjects for future work.

Such integration of PDEs with ML opens a door for a tighter data-knowledge connection, which may potentially influence the further development of the physics-based modelling with ML for data-driven thermal fluid models.

## NOMENCLATURE

| | | | |
|---|---|---|---|
| **Mathematical symbols** | | | |
| $\Delta t$ | Time step size, s | $L$ | Dimensions of the domain, m |
| $\varepsilon$ | Tolerance for convergence | | Number of layers in a NN |
| $\phi$ | An arbitrary field | $m$ | Mesh nodes number in $y$-direction |
| $\varphi$ | Resulting field after convolution of $\phi$ | $N_{data}$ | Number of datasets |
| $\eta$ | Learning rate | $n$ | Mesh nodes number in $x$-direction |
| $\mu$ | Dynamic viscosity, Pa·s | $p$ | Pressure, Pa |
| $\rho$ | Density, kg/m$^3$ | $T$ | Temperature, K |
| $\sigma$ | Sigmoid activation function | $t$ | Time, s |
| $A$ | Advection term, m/s$^2$ | $u$ | $x$-velocity, m/s |
| $b$ | Biases of a NN | $v$ | $y$-velocity, m/s |
| $C$ | Cost (loss) function | $vel$ | Absolute velocity, m/s |
| $c$ | Constant | $W$ | Weights of a NN |
| $D$ | Diffusion term, kg/(m$^2$·s$^2$) | $x$ | Horizontal coordinate, m |
| $h$ | Mesh size, m | $y$ | Vertical coordinate, m |
| $k$ | Kernel (filter) | | |
| | Thermal conductivity, W/(m·K) | | |
| **Subscripts** | | | |
| 0 | Initial value | $out$ | Output value |
| $func$ | Exact functional value | $p$ | Pressure |
| $i$ | Mesh node's $x$-index | $sol$ | Solution |
| $inp$ | Input value | $targ$ | Target value |
| $j$ | Mesh node's $y$-index | $vel$ | Velocity |
| $h$ | Finite volume approximation | $w$ | *Wall* |
| $max$ | Maximum value | $x$ | $x$-projection of a vector |
| $nn$ | NN-based value | $y$ | $y$-projection of a vector |
| **Superscripts** | | | |
| * | Predictor step (preliminary) value | $T$ | Transposed tensor |
| $n$ | Time step number | | |
| **Dimensionless Numbers** | | | |
| Re | Reynolds number | CFL | Courant-Friedrichs-Lewy number |
| **Acronyms** | | | |
| ANN | Artificial Neural Network | ML | Machine Learning |
| API | Application Programming Interface | NN | Neural Network |
| CG | Coarse-Grid | PDE | Partial Differential Equation |
| CFD | Computational Fluid Dynamics | QoI | Quantity of Interest |
| CNN | Convolutional Neural Network | RANS | Reynolds-Averaged Navier-Stokes |
| CR | Closure Relation (or closure model) | RMSE | Root Mean Square Error |
| DD | Data-Driven | SET | Separate-Effect Test |
| DFNN | Deep Feedforward Neural Network | SS | Steady State |
| DL | Deep Learning | VV&UQ | Verification, Validation and Uncertainty Quantification |
| DNS | Direct Numerical Simulations | | |

# 1. INTRODUCTION

The field of fluid dynamics is rapidly advancing and generates a large amount of data from simulations and experiments at multiple spatiotemporal scales. Traditionally, the fluid dynamics modelling is being performed using the physics (knowledge) base design (e.g. the Navier-Stokes equations or Lattice Boltzmann methods), which allowed to make a great progress in science and engineering in the past centuries. However, there are serious challenges associated with current equations-based analysis of fluids, including high dimensionality and nonlinearity, which defy closed-form solutions and limit real-time optimization and control efforts [1]. Current situation may be characterized as "data-rich, knowledge-poor": the complexity of underlying physics is so high that it is almost impossible to learn and extract new knowledge from a large amount of the observed data directly [2]. Machine learning (ML) offers a plethora of different techniques to extract information from this data, which then can be potentially translated into knowledge about the underlying physics. Thus, ML algorithms can augment domain knowledge and automate tasks related to flow control and optimization and provides a powerful information processing framework that can augment, and possibly even transform, current lines of fluid dynamics research and industrial applications [1].

One of the application fields of fluid dynamics is nuclear thermal hydraulics. Paper [3] suggests a classification of ML frameworks for data-driven (DD) thermal fluid models based on the requirements of data and knowledge; the classification briefly summarized in Table 1.1. The authors [1] distinguished 5 different types of ML frameworks employed in the literature for thermal fluid models:

- Type 1 (Physics-Separated) ML. Perhaps it is the most widespread one and it is usually employed when there is a necessity to build a surrogate closure relation (CR) using data obtained in a separate-effect[1] test (SET) and later use it. The main assumption here is that the conservation equations (usually in PDE form) and CRs (which close the PDEs), are scale separable. This framework requires a substantial knowledge base about the physics, but is not as data-hungry as the other ML frameworks. Possible examples include prediction of the minimum film boiling temperature [4], critical flow prediction [5], and many others [3].

[1] Separate effects describe the behavior of single component or characteristics of one phenomenon.

- Type 2 (Physics-Evaluated) ML is employed when high-fidelity data are used to reduce uncertainties and/or errors in low-fidelity simulations. In other words, it is a DD approach to use high-fidelity codes to calibrate low-fidelity codes [6]. This framework requires the availability of a sufficient amount of high-fidelity data (or other information about a solution with higher accuracy), which poses a restriction for real engineering applications (e.g. Direct Numerical Simulations (DNS) are still very computationally expensive), and substantial knowledge base for choosing appropriate CRs (if applicable) to inform the low-fidelity simulations (e.g. an appropriate turbulence model for the Reynolds-Averaged Navier-Stokes (RANS) simulations). Hanna B. et al. [7] used high-fidelity simulations to inform Coarse-Grid Computational Fluid Dynamics (CG-CFD) simulations using artificial NNs (ANNs) and random forests. It is also interesting to mention a paper [8], where the authors suggested a framework to predict errors in CG simulations without high-fidelity data: instead, they use an adjoint-based error estimation method. Perhaps, such approach has a limited range of applications since the adjoint-based methods require linearization of the Quantities of Interests (QoIs), which is only possible for a limited range of problems [9].
- Type 3 (Physics-Integrated) ML is employed when there is a need to build a CRs without having a target data for direct outputs from a NN; in this case the scale separation assumption is not required [3] and a CR is trained together with the solution of PDEs. This framework is firstly introduced in [3] and its performance is demonstrated on 2D heat conduction equation (see Section 3.1): CR for non-constant temperature-dependent thermal conductivity is built using ANNs, which is integrated with 2D heat conduction PDE. Possibly similar framework is proposed in [10, 11], where the authors are used a NN integrated with the Navier-Stokes equations to predict velocity and pressure fields using a scalar quantity fields; however, the PDEs are whether encoded in a NN, or solved using automatic differentiation available through ML libraries. At the same time, the authors [11] claim that “the proposed methods should not be viewed as replacements of classical numerical methods for solving PDEs … Such methods have matured over the last 50 years and, in many cases, meet the robustness and computational efficiency standards required in practice.” Current work is aimed at further investigations of the Type 3 ML framework for its applicability for the embedding (or integration) of a NN in the numerical solution of the Navier-Stokes equations (see Section 3.2). Thus, it can be viewed as a “bridge” between ML and classical numerical methods.

- Type 4 (Physics-Recovered) ML is aimed at recovering the exact form of PDEs by finding optimal candidates for unknown terms (e.g. advection, diffusion terms, etc.). The knowledge of possible candidates is necessary to provide a ML algorithm with a list of these terms. One of the examples is a usage of a genetic algorithm to choose best candidates for unknown terms in PDEs [12].
- Type 5 (Physics-Discovered) ML is a very data-hungry framework since no knowledge base is available: the DD-framework is aimed at recovering PDEs describing the physics (or their solution) from big data [13]. For example, Guo X. et al. [14] employed Convolutional Neural Networks (CNNs) for steady flows approximation; Raiisi M. et al. [15] proposed a framework to learn PDEs from small data.

ML for fluid dynamics is a rapidly advancing area. At the time when the classification [3] into Type 1, …, Type 5 was proposed, these different types were not intersecting, However, the advances show that there is a possibility to combine them: Type 5 (learning of PDEs) might be used to create a physics-informed framework (which is a variation of a Type 3) [10, 11]; Type 3 might be potentially applicable to predict errors in low-fidelity simulations (Type 2). Therefore, the classification [3] is very useful for understanding of different approaches employed for DD closures, not obsolete, but requires further adjustments.

Recent comprehensive review of ML application in fluid mechanics is performed in [1]. The authors analyzed different ML tasks (supervised, semi-supervised, and unsupervised learning and further sub-branches) and their possible applications for flow modelling and flow optimization and control. A brief summary is presented in Table 1.2 with some examples found in the literature. Some of the frameworks in Table 1.2 are classified according to Table 1.1, while the others were not considered in [3] (denoted by N/A).

ML is a fast-growing field and the researchers are continuously trying to tackle challenges for its employment in fluid dynamics that are also outlined in the review [1]. These challenges are not specific only for fluid dynamics field, but rather a more general problems of artificial intelligence (a reader may get further information reading the references provided after each of them):

- ML algorithms often come without guarantees for performance, robustness, or convergence, even for the well-defined tasks. How can interpretability, generalizability, and explainability of the results be achieved? [27, 28]

- Incorporating and enforcing known flow physics is a challenge and opportunity for ML algorithms. Can we hybridize DD and first principle approaches in fluid dynamics? [10, 29]
- Meta-learning (learning to learn) – how can we make ML algorithms continuously learning? [30, 31]
- Transfer learning – how to incorporate previous knowledge into a new ML algorithm? [32]
- How to improve the performance of the ML for extrapolation mode and assess data coverage condition? [33]
- How to perform Verification, Validation and Uncertainty Quantification (VV&UQ) of ML and DD frameworks? [34, 35]

Unfortunately, pure DD approaches often ignore knowledge base that was generated throughout centuries. A paper [29] is a comprehensive survey of the current attempts in the ML-community to integrate physics-based modeling with ML, which is "fundamentally different from mainstream practices in the ML community for making use of domain-specific knowledge" [29]. Some of the objectives of the physics-ML integration are: improving predictions of physical models, reducing the dimensionality and order of physics-based models, parametrization, inverse modelling, forward solving of PDEs, discovering governing equations, data generation, UQ. The methods reviewed include introduction of the physics-guided cost function (when additional terms are introduced to the cost function to penalize physically inconsistent results); physics-guided initialization (when initial parameters such as weights and biases of a NN are initialized with usage of some previous knowledge, which is tightly connected with transfer learning); physics-guided design of architecture (when the architecture of a NN is designed for a specific task according to the knowledge-base); residual modelling (error prediction, or Type 2 ML); hybrid physics-ML models (when ML-models are integrated with physics-based model to improve the predictions or compensate the lack of data, which is, perhaps, a more general concept for Type 3 ML).

From the framework development point of view, knowledge[2] may be implemented in ML on a pre-training, training, and post-training stages [36]. Pre-training phase may include implementation of a specific architecture for a NN, which will be knowledge-informed (e.g. Muralidhar N. et al. [37] used physics-guided design to improve prediction of the drag force). On the training phase, physics-informed cost function [10], or transfer learning [32] techniques may be used. Finally, data-refinement [38] is a tool to use knowledge to inform DD-models on a post-training phase.

[2] Knowledge, physics and PDEs are used as synonyms in this paper. It is the authors' believe that definitions for "physics", "knowledge", "physics-integration", "physics-informed", and other terms widely used in ML community, require further refinement and clarification.

The aim of this paper is to investigate a possibility to integrate ML with first principle approaches in fluid dynamics. specifically, Type 3 framework for the Navier-Stokes equations (2D lid-driven cavity with non-constant velocity-dependent dynamic viscosity is considered; a surrogate for the dynamic viscosity is built using a deep feedforward NN (DFNN)). Section 2 provides the information on the numerical solution methodology of the Navier-Stokes equations using the Tensorflow library [39] for Python programming language [41]; Section 3 is describing Type 3 ML framework and its architecture; Section 4 consists of the case study formulation and solution verification; Section 5 is devoted to the discussion of the results and setting tasks for further investigations.

Table 1.1. Classification of ML frameworks for DD thermal fluid models [3].

| | **Type 1**<br>**Physics-Separated ML** | **Type 2**<br>**Physics-Evaluated ML** | **Type 3**<br>**Physics-Integrated ML** | **Type 4**<br>**Physics-Recovered ML** | **Type 5**<br>**Physics-Discovered ML** |
|---|---|---|---|---|---|
| | **Regression Task** | | | | |
| Goal | Develop CRs by using SET data | Reduce the uncertainty for conservation equations by utilizing high-fidelity data to inform low-fidelity simulations | Develop CRs without scale separation assumption and target data for a NN | Recover the exact form of PDEs | Recover the PDEs or their solution from data |
| Major assumption | Conservation equations and CRs are scale separable<br>Knowledge base for physics is substantial | High-fidelity data are available<br>Knowledge base for CRs selection is substantial | PDE for physics is available | Candidates for terms in PDEs are available | Knowledge of physics is unavailable<br>Abundance of training data |

Table 1.2. ML for fluid mechanics [1].

| | **Flow modelling with ML** | | | | | | | |
|---|---|---|---|---|---|---|---|---|
| | **Flow feature extraction** | | | | **Modelling flow dynamics** | | | |
| | **Dimensionality reduction** | **Sparse and randomized methods** | **Super-resolution and flow cleaning** | **Clustering and classification** | **Linear models through nonlinear embeddings** | **NN modelling** | **Parsimonious nonlinear models** | **CRs with ML** |
| | **Regression Task (mostly)** | | | **Classification Task** | **Regression Task** | | | |
| Goal | Simplify data and find low dimensional feature spaces and more effective non-linear coordinates | Decrease the amount of data | Improve simulations and experiments | Classify different (physical) domains in a problem | Identify a coordinate system where nonlinear dynamics appear linear | Uncover latent variables and relations from data | Construct a simple model for balancing predictive accuracy and complexity | Build CRs for conservation equations |
| Example | Reconstruct near wall velocity field in a turbulent channel flow using wall pressure and shear [16] | Decrease the amount of data for compact representations of wall-bounded turbulence [17] | Turbulence reconstruction inside a cell for large eddy simulations (LES) [18] | Two-phase flow regime classification [19] | Extraction of spatiotemporal coherent structures from time series data of fluid flows [20] | Learn PDEs or their solutions [21] | Conservation equations reconstruction using genetic algorithms [22] | Reynolds stress tensor for RANS [23] |
| Type [1] | N/A | N/A | 2 | N/A | N/A | 5 | 4 | 1 |

Table 1.2. ML for fluid mechanics [1] (continuation).

| | **Flow optimization and control using ML** | | |
|---|---|---|---|
| | **NNs for control** | **Genetic algorithms for control** | **Flow control via reinforcement learning** |
| | **Regression Task (mostly)** | | |
| Goal | Optimization and control of the flow parameters | | |
| Example | Turbulent flow control/drag optimization [24] | Shear flow control GA [25] | Reproduce the dynamics of hydrological systems [26] |
| Type [1] | N/A | N/A | N/A |

## 2. Governing Equations and their Numerical Solution

### 2.1. The Navier-Stokes Equations Discretization

In this paper the incompressible 2D Navier-Stokes equations with non-constant dynamic viscosity (2.1.1) and (2.1.2) are solved using the Chorin's projection method [42] and end-to-end open source ML platform Tensorflow developed by Google (API version 1.15) [39] on the Python 3.7 programming language [41]. Additionally, due to a relatively slow solution on the Python 3.7, analogous Fortran program is developed for fast data generation and manipulation (see Section 4.3).

Thus, the governing equations are

$$\nabla \cdot \vec{u} = 0 \tag{2.1.1}$$

$$\frac{\partial \vec{u}}{\partial t} + \nabla\left(\vec{u}\vec{u}^T\right) = -\frac{1}{\rho}\nabla p + \frac{1}{\rho}\nabla \cdot \left[\mu\left(\nabla \vec{u} + \nabla \vec{u}^T\right)\right] \tag{2.1.2}$$

where $\vec{u} = \vec{u}\left(t, x, y\right)$ is 2D velocity vector; $t$ is time; ρ is density; $p = p\left(t, x, y\right)$ is pressure; μ is non-constant dynamic viscosity.

The Navier-Stokes equations (2.1.1) and (2.1.2) are discretized using the finite-volume method on the staggered uniform grid with characteristic size of a control volume $h$ (Fig. 2.1.1):

$$\begin{aligned} u^{*}_{i+1/2,j} &= u^{n}_{i+1/2,j} + \Delta t\left[-\left(A_x\right)^{n}_{i+1/2,j} + \frac{1}{\rho}\left(D_x\right)^{n}_{i+1/2,j}\right] \\ v^{*}_{i,j+1/2} &= v^{n}_{i,j+1/2} + \Delta t\left[-\left(A_y\right)^{n}_{i,j+1/2} + \frac{1}{\rho}\left(D_y\right)^{n}_{i,j+1/2}\right] \end{aligned} \tag{2.1.3}$$

$$\begin{gathered} \frac{1}{\rho}\nabla_h^{\,2} p^{n+1} = \frac{1}{\Delta t}\nabla_h \cdot \vec{u}^{*} \\ \frac{p^{n+1}_{i+1,j} + p^{n+1}_{i-1,j} + p^{n+1}_{i,j+1} + p^{n+1}_{i,j-1} - 4p^{n+1}_{i,j}}{h^2} = \frac{\rho}{\Delta t}\left(\frac{u^{*}_{i+1/2,j} - u^{*}_{i-1/2,j} + v^{*}_{i,j+1/2} - v^{*}_{i,j-1/2}}{h}\right) \end{gathered} \tag{2.1.4}$$

$$\begin{aligned} u^{n+1}_{i+1/2,j} &= u^{*}_{i+1/2,j} - \frac{\Delta t}{\rho h}\left(p^{n+1}_{i+1,j} - p^{n+1}_{i,j}\right) \\ v^{n+1}_{i,j+1/2} &= v^{*}_{i,j+1/2} - \frac{\Delta t}{\rho h}\left(p^{n+1}_{i,j+1} - p^{n+1}_{i,j}\right) \end{aligned} \tag{2.1.5}$$

where $n$ denotes the number of a time step; $u$ and $v$ are $x$- and $y$-velocities, respectively; $\Delta t$ is time step size; $A$ and $D$ are advection and diffusion terms, respectively; $\nabla_h$ and $\nabla_h^{\,2}$ denote numerical approximations for gradient and Laplacian operators, respectively; superscript * denotes preliminary value for velocities (predictor step).

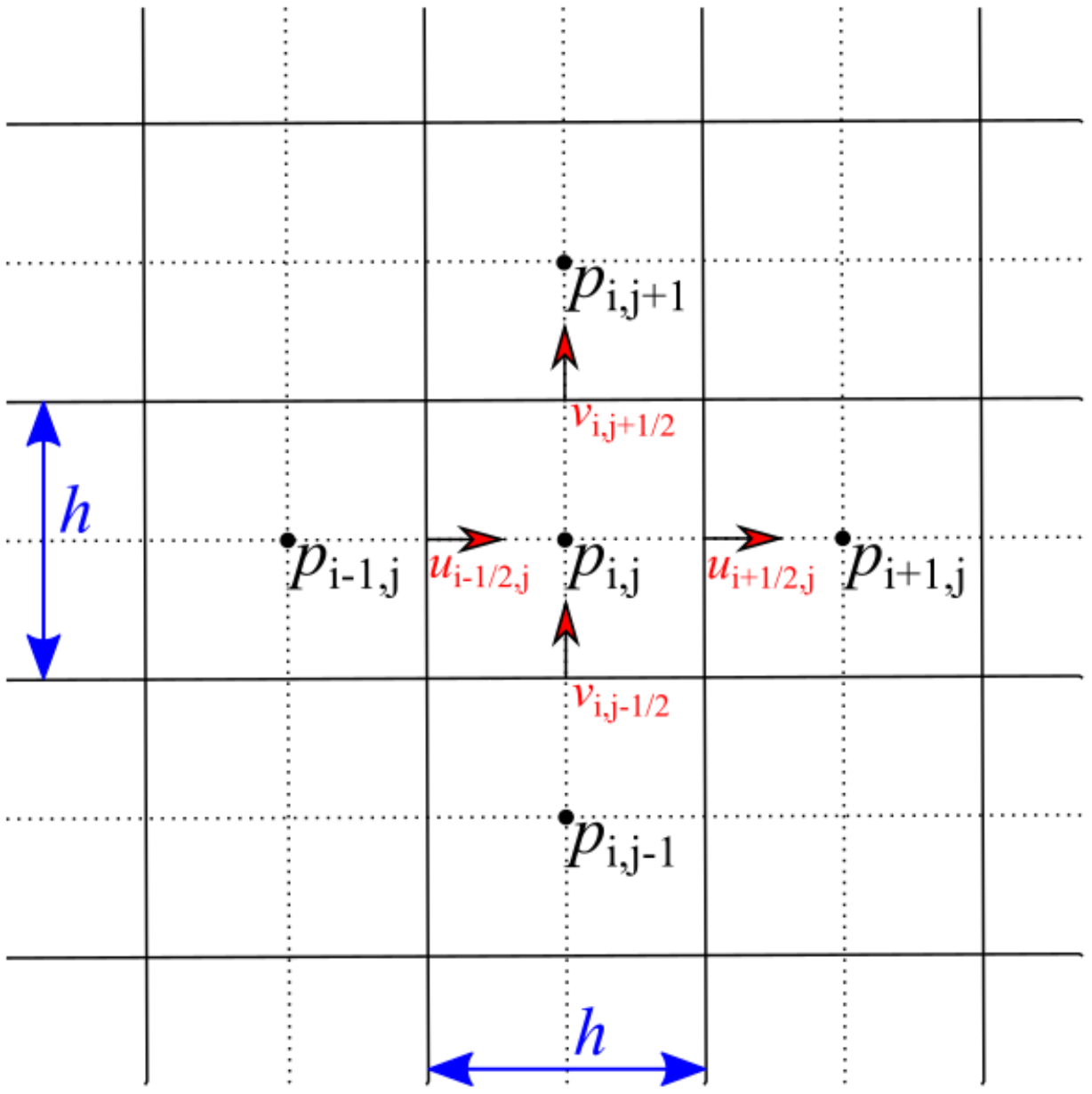

Fig. 2.1.1. Staggered uniform grid: pressure, density, and dynamic viscosity are in the centers of the control volumes; velocities are on the edges.

To ensure the numerical stability of the scheme, the Courant-Friedrichs-Lewy $CFL = u_w \cdot \Delta t / h$ number was set 0.25 for cases with constant dynamic viscosity (solution verification, Section 4.2) and 0.05 for cases with non-constant dynamic viscosity. Here $vel_{max}$ is the maximum absolute velocity in the domain.

The solution procedure of the above equations is known as the Chorin's projection method [42], which consists of prediction step (2.1.3) to find preliminary velocities $u^*$ and $v^*$ (ignoring pressure gradient); pressure-Poisson equation (2.1.4), which has to be iteratively solved on each time step; and projection step (2.1.5) to find the velocities on a new time step. This method is very effective and allows splitting velocity and pressure and solve for them independent sets of equations. In this study only steady state (SS) solutions are considered: as the SS condition the value $\varepsilon_{vel} = \max(|u^n - u^{n+1}|, |v^n - v^{n+1}|) = 10^{-8}$ m/s is adopted as a convergence criterion. At the same time, the pressure equation iterations are limited to 100, or until a convergence $\varepsilon_p = \max|p^n - p^{n+1}| = 10^{-8}$ Pa is reached.

The discretizations of the advection and diffusion terms are obtained using the finite volume method [44]:

$$\left(A_x\right)^n_{i+1/2,j} = \frac{1}{4h}\Big\{\left(u^n_{i+3/2,j} + u^n_{i+1/2,j}\right)^2 - \left(u^n_{i+1/2,j} + u^n_{i-1/2,j}\right)^2 +$$
$$+\left(u^n_{i+1/2,j+1} + u^n_{i+1/2,j}\right)\left(v^n_{i+1,j+1/2} + v^n_{i,j+1/2}\right) - \left(u^n_{i+1/2,j} + u^n_{i+1/2,j-1}\right)\left(v^n_{i+1,j-1/2} + v^n_{i,j-1/2}\right)\Big\}$$
$$\left(A_y\right)^n_{i,j+1/2} = \frac{1}{4h}\Big\{\left(v^n_{i,j+3/2} + v^n_{i,j+1/2}\right)^2 - \left(v^n_{i,j+1/2} + v^n_{i,j-1/2}\right)^2 +$$
$$+\left(v^n_{i+1,j+1/2} + v^n_{i,j+1/2}\right)\left(u^n_{i+1/2,j+1} + u^n_{i+1/2,j}\right) - \left(v^n_{i,j+1/2} + v^n_{i-1,j+1/2}\right)\left(u^n_{i-1/2,j+1} + u^n_{i-1/2,j}\right)\Big\} \tag{2.1.6}$$

$$\left(D_x\right)^n_{i+1/2,j} = \frac{1}{h^2}\Big[2\mu_{i+1,j}\left(u^n_{i+3/2,j} - u^n_{i+1/2,j}\right) - 2\mu_{i,j}\left(u^n_{i+1/2,j} - u^n_{i-1/2,j}\right) +$$
$$+\mu_{i+1/2,j+1/2}\left(u^n_{i+1/2,j+1} - u^n_{i+1/2,j} + v^n_{i+1,j+1/2} - v^n_{i,j+1/2}\right) -$$
$$-\mu_{i+1/2,j-1/2}\left(u^n_{i+1/2,j} - u^n_{i+1/2,j-1} + v^n_{i+1,j-1/2} - v^n_{i,j-1/2}\right)\Big]$$
$$\left(D_y\right)^n_{i,j+1/2} = \frac{1}{h^2}\Big[2\mu_{i,j+1}\left(v^n_{i,j+3/2} - v^n_{i,j+1/2}\right) - 2\mu_{i,j}\left(v^n_{i,j+1/2} - v^n_{i,j-1/2}\right) +$$
$$+\mu_{i+1/2,j+1/2}\left(v^n_{i+1,j+1/2} - v^n_{i,j+1/2} + u^n_{i+1/2,j+1} - u^n_{i+1/2,j}\right) -$$
$$-\mu_{i-1/2,j+1/2}\left(v^n_{i,j+1/2} - v^n_{i-1,j+1/2} + u^n_{i-1/2,j+1} - u^n_{i-1/2,j}\right)\Big] \tag{2.1.7}$$

The values of dynamic viscosity on the edges of the control volumes are found using the simple averaging, e.g.

$$\mu_{i+1/2,j+1/2} = \frac{1}{4}\left(\mu_{i,j} + \mu_{i+1,j} + \mu_{i,j+1} + \mu_{i+1,j+1}\right) \tag{2.1.8}$$

As it could be seen, the adopted time and space discretization schemes are the simplest ones and might be unstable for high Re numbers, but in this study, they are used for simplicity and demonstration purpose. They are provided in this paper to help a reader better understand the solution methodology of these equations using the Tensorflow library, which is described in the next section.

### 2.2. Solution of the Navier-Stokes Equations using the Tensorflow Library

Equations (2.1.3) – (2.1.8) are solved using the Tensorflow ML library [39] employing the convolutional layer model *tf.nn.conv2d* [43]. Before diving into the solution procedure, we discuss basics of the convolutional operations in the Tensorflow (and other ML libraries).

Convolution operations are used in deep learning (DL) to design a special class of ANNs called Convolutional NNs (CNNs). The CNNs are ubiquitous in DL and they have shown big advantages comparing to the DFNNs for such problems as image recognition, computer vision, text classification, etc. [45], where data have a spatial structure.

Fig. 2.2.1 demonstrates how a convolution operation is applied to a 2D scalar field $\phi$. The kernel (or filter) $k_{i,j}$ is moving along two dimensions: from left to right and from top to bottom performing the convolution of the field $\phi$ over the region where it is applied. As a result, a new scalar field $\varphi$ is obtained, which is calculated using a following rule:

$$\begin{aligned}\varphi_{i-1,j+1} = k_{1,1}\phi_{i-2,j+2} + k_{2,1}\phi_{i-1,j+2} + k_{3,1}\phi_{i,j+2} + \\ + k_{1,2}\phi_{i-2,j+1} + k_{2,2}\phi_{i-1,j+1} + k_{3,2}\phi_{i,j+1} + \\ + k_{1,3}\phi_{i-2,j} + k_{2,3}\phi_{i-1,j} + k_{3,3}\phi_{i,j}\end{aligned} \qquad (2.2.1)$$

This new scalar field $\varphi$ has a lower dimension than $\phi$ depending on the size of a kernel (not necessarily). Please note that the explained convolution is the simplest one: there are many different opportunities for a user to specify parameters of a convolution (e.g. padding, stride, dilation, etc. [43]), but such basic understanding is enough to understand how discretized PDEs can be solved using the convolution.

In ML, when CNNs are employed, the kernel values $k_{i,j}$ are being trained to extract features from the input fields in the same manner as weights in a DFNN, but it is also possible to set them manually (without further changes). For instance, applying a pre-set kernel

$$\ker_1 = \begin{bmatrix} 0 & 1 & 0 \\ 1 & 0 & 1 \\ 0 & 1 & 0 \end{bmatrix} \qquad (2.2.2)$$

to a pressure field, one will obtain a new field with values:

$$\varphi_{i,j} = p_{i+1,j}^{n+1} + p_{i-1,j}^{n+1} + p_{i,j+1}^{n+1} + p_{i,j-1}^{n+1}$$

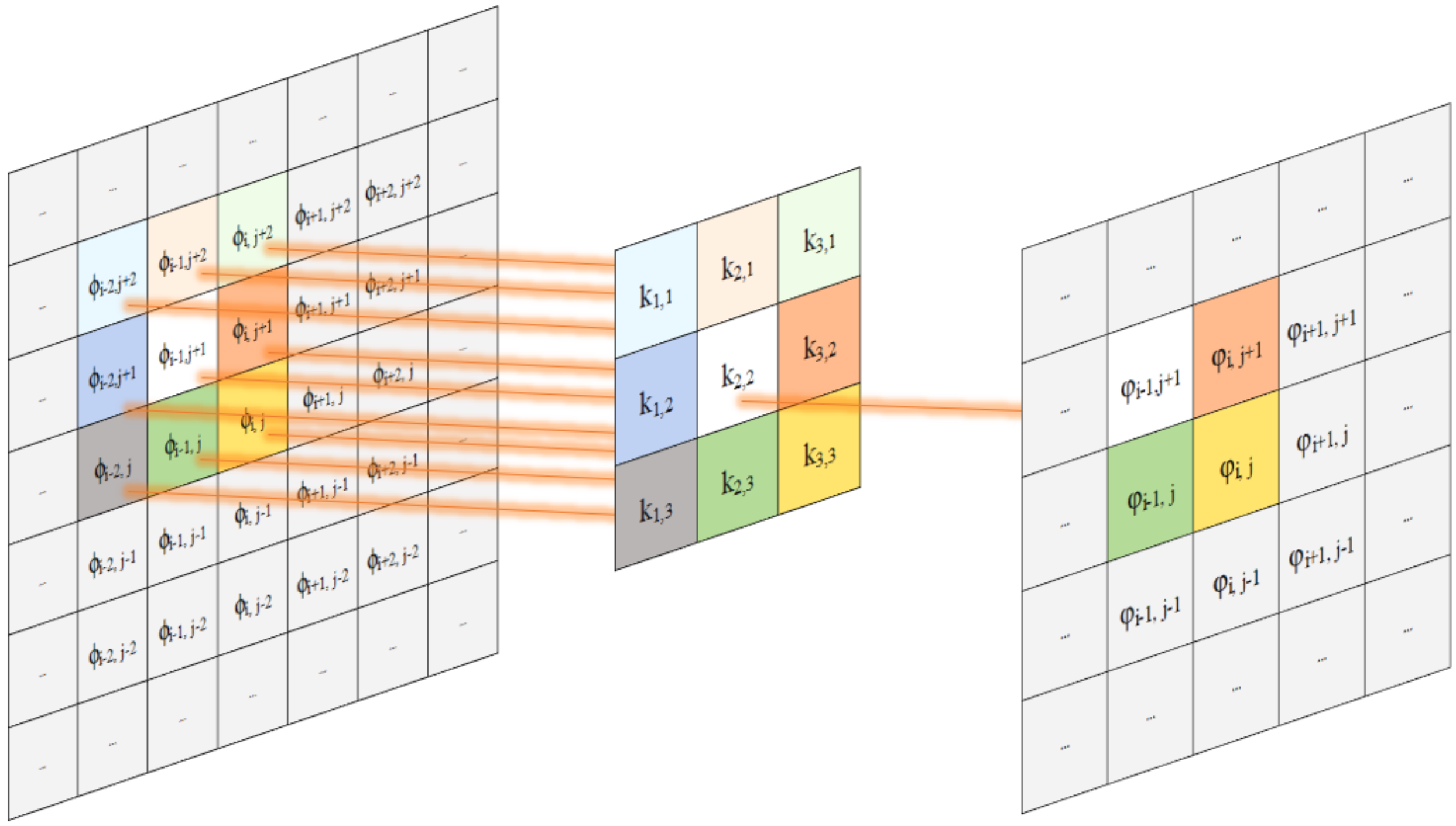

Fig. 2.2.1. Convolution operation applied for a 2D field ϕ ($m$×$n$) using a kernel $k$ (3×3) to obtain a new field φ (($m$ – 2)×($n$ – 2)).

To summarize this discussion, we provide the kernels and solution procedure of the Navier-Stokes equations (2.1.3) – (2.1.8) below:

$$\mathrm{ker}_1 = \begin{bmatrix} 0 & 1 & 0 \\ 1 & 0 & 1 \\ 0 & 1 & 0 \end{bmatrix} \qquad \mathrm{ker}_4 = \begin{bmatrix} 0.25 & 0.25 \\ 0.25 & 0.25 \end{bmatrix}$$

$$\mathrm{ker}_2 = \begin{bmatrix} -1 & 1 \end{bmatrix} \qquad \mathrm{ker}_5 = \begin{bmatrix} 1 & 1 \end{bmatrix} \tag{2.2.3}$$

$$\mathrm{ker}_3 = \begin{bmatrix} -1 \\ 1 \end{bmatrix} \qquad \mathrm{ker}_6 = \begin{bmatrix} 1 \\ 1 \end{bmatrix}$$

$$\left(A_x\right)^n = \frac{1}{4h}\left\{\mathrm{ker}_2\left[\mathrm{ker}_5\left(u^n\right)\cdot\mathrm{ker}_5\left(u^n\right)\right]+\mathrm{ker}_3\left[\mathrm{ker}_6\left(u^n\right)\cdot\mathrm{ker}_5\left(v^n\right)\right]\right\}$$

$$\left(A_y\right)^n = \frac{1}{4h}\left\{\mathrm{ker}_3\left[\mathrm{ker}_6\left(v^n\right)\cdot\mathrm{ker}_6\left(v^n\right)\right]+\mathrm{ker}_2\left[\mathrm{ker}_6\left(u^n\right)\cdot\mathrm{ker}_5\left(v^n\right)\right]\right\} \tag{2.2.4}$$

$$\left(D_x\right)^n = \frac{1}{h^2}\left\{2\cdot\mathrm{ker}_2\left[\mu^n\cdot\mathrm{ker}_2\left(u^n\right)\right]+\mathrm{ker}_3\left[\mathrm{ker}_4\left(\mu^n\right)\cdot\left(\mathrm{ker}_3\left(u^n\right)+\mathrm{ker}_2\left(v^n\right)\right)\right]\right\}$$

$$\left(D_y\right)^n = \frac{1}{h^2}\left\{2\cdot \mathrm{ker}_3\left[\mu^n \cdot \mathrm{ker}_3\left(v^n\right)\right] + \mathrm{ker}_2\left[\mathrm{ker}_4\left(\mu^n\right)\cdot\left(\mathrm{ker}_2\left(v^n\right)+\mathrm{ker}_3\left(u^n\right)\right)\right]\right\}$$

$$u^* = u^n + \Delta t\left[-\left(A_x\right)^n + \frac{1}{\rho}\left(D_x\right)^n\right]$$

$$v^* = v^n + \Delta t\left[-\left(A_y\right)^n + \frac{1}{\rho}\left(D_y\right)^n\right]$$

$$p^{n+1} = \frac{1}{4}\left\{\mathrm{ker}_1\left(p\right) - \frac{\rho h}{\Delta t}\left[\mathrm{ker}_2\left(u^*\right)+\mathrm{ker}_3\left(v^*\right)\right]\right\} \tag{2.2.5}$$

$$u^{n+1} = u^* - \frac{\Delta t}{\rho h}\mathrm{ker}_2\left(p^{n+1}\right)$$

$$v^{n+1} = v^* - \frac{\Delta t}{\rho h}\mathrm{ker}_3\left(p^{n+1}\right) \tag{2.2.6}$$

A tricky part is the setting boundary conditions in the Tensorflow solution. Do to it, symmetric padding operation *tf.pad* was used for pressure and dynamic viscosity; zero padding *tf.pad*, dimension expanding *tf.expand_dims*, and tensor concatenation *tf.concat* operations were used for velocities.

Note that the Tensorflow performs tensor operations, unlike Fortran which calculates the values element by element using indexes $i$ and $j$, and, therefore, the values in Eqs. (2.2.4) – (2.2.6) are tensors. It gives some advantages and disadvantages that are discussed in Section 4.3.

## 3. TYPE 3 MACHINE LEARNING: EMBEDDING A NEURAL NETWORK IN PARTIAL DIFFERENTIAL EQUATIONS

### 3.1. Basic Introduction to Type 3 Machine Learning

The Type 3 ML framework is firstly proposed in the paper [3] and its performance is demonstrated on stationary 2D heat conduction PDE with non-constant temperature-dependent thermal conductivity $k$:

$$\frac{\partial}{\partial x}\left[k\left(T\right)\frac{\partial T}{\partial x}\right] + \frac{\partial}{\partial y}\left[k\left(T\right)\frac{\partial T}{\partial y}\right] = 0 \tag{3.1.1}$$

where $T$ is temperature, $x$ and $y$ are coordinates.

The functional dependence for the thermal conductivity considered is the normal (Gauss) distribution

$$k(T)=\frac{c_1}{c_2\sqrt{2\pi}}e^{-\frac{(T-c_3)^2}{2c_2^2}} \tag{3.1.2}$$

where $c_1$, $c_2$, $c_3$ are some fixed parameters.

The heat conduction equation (3.1.1) might be discretized as

$$T_{i,j}^{n+1}=\frac{k_{i+1/2,j}T_{i+1,j}+k_{i,j+1/2}T_{i,j+1}+k_{i-1/2,j}T_{i-1,j}+k_{i,j-1/2}T_{i,j-1}}{k_{i-1/2,j}+k_{i,j-1/2}+k_{i+1/2,j}+k_{i,j+1/2}} \tag{3.1.3}$$

and solved on the staggered grid iteratively using the Jacobi iterations with the Tensorflow library [39] employing a similar technique as described in Section 2.2 (this case is much simpler than the Navier-Stokes equations).

A NN is embedded in the solution of the heat conduction equation as it shown in Fig. 3.1.1.

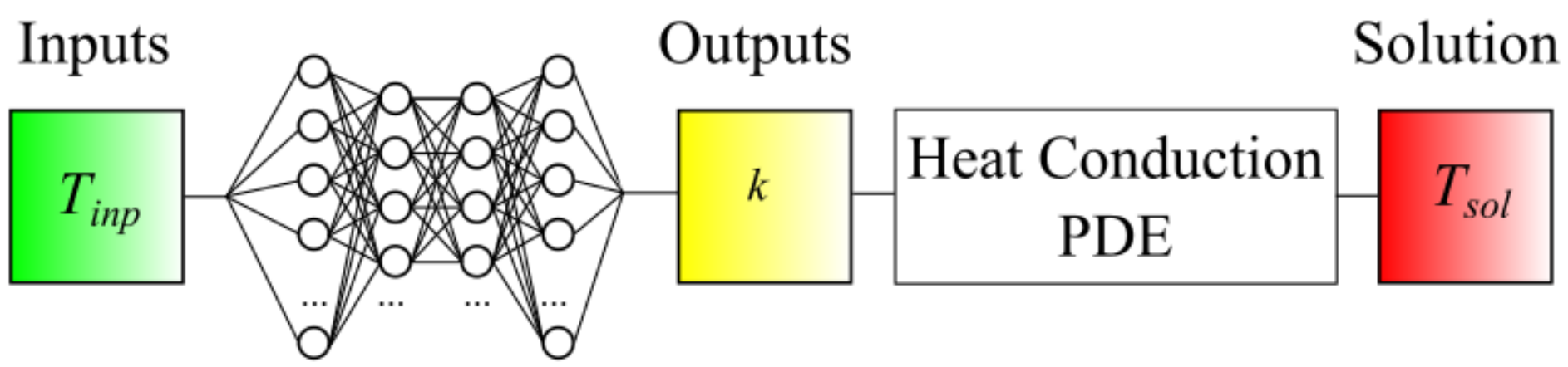

Fig. 3.1.1. Type 3 ML: a NN is embedded into the solution of the heat conduction equation [3].

Note, that for the Type 3 ML it is crucial to solve the PDE(s) using the same library (e.g. Tensorflow [39], Pytorch [40]) as the NN is based on, because the Python script must have a traceable connection between the solution of the PDE ($T_{sol}$) and outputs from the NN ($k$). At the same time, one need not to have the thermal conductivity values as training targets at all: the cost function uses only the solution of the PDE $T_{sol}$ and target values of temperatures $T_{targ}$:

$$C=\frac{1}{2}\frac{\sum\left(T_{sol_{i,j}}-T_{targ_{i,j}}\right)^2}{N_{data}} \tag{3.1.4}$$

where $N_{data}$ is the number of training datasets, and $T_{sol_{i,j}}=T\left(k_{nn}\right)$.

The authors [3] proposed a following algorithm for Type 3 framework implementation (Fig. 3.1.2). Firstly, it is necessary to prepare the input to the NN (initial guess for the temperature distribution $T_{inp}^{0}$); these values are updated after each global iteration. The training target temperatures $T_{targ}$ might be obtained in the experiments, or in calculations; these values are fixed and do not change. Since we maintain the connection between the $T_{inp}$ and $T_{sol}$ via the NN and numerical solution of the PDE (Fig. 3.1.1), it is possible to train the NN by updating its weights and biases without values of the thermal conductivity $k$ using the cost function Eq. (3.1.4). When necessary number of iterations is performed (or desired value of the cost function (accuracy) is reached), the PDE is separately solved using the NN-based thermal conductivity and new values $T_{sol}$ are obtained. Then these values are used to update $T_{inp}$ and we proceed to a next global iteration, which are executed until some ad-hoc criteria are met.

Note, that this framework is very time consuming since it includes several iteration loops (global iterations, iterations to train a NN, iterations to solve a PDE), but at the same time it promises a high-potential impact in complex thermal fluid problems where the separation of scales or physics decomposition may involve significant errors [3].

| |
|---|
| **Prepare input**: Initial guess for the temperature field $T_{inp} = T_{inp}^{0}$ |
| **Prepare targets**: Temperature field from the experiments (or simulations) $T_{targ}$ |
| **For** *it* **in range** (0, $it_{max}$) **do** {global iterations} |
| **For** *epoch* **in range** (0, $epoch_{max}$) **do** {iterations to train the NN} |
| Train the NN (to build a surrogate for $k(T_{inp})$) to minimize $C$ (Eq. 3.1.1) |
| Solve the PDE (3.1.1) using the NN-based $k(T_{inp})$ and obtain $T_{sol}$ {Jacobi iterations} |
| Replace $T_{inp}$ by $T_{sol}$ and proceed to the next global iteration |

Fig. 3.1.2. Proposed algorithm for Type 3 ML in [3].

In the next section a Type 3 ML framework for training a surrogate model of the velocity-dependent dynamic viscosity inside the Navier-Stokes equations is described. At the same time, we note that it is not necessary to constantly update the $T_{inp}$ values if SS solutions are considered. Indeed, considering a following limit

$$\lim_{it \to \infty} T_{inp} = T_{targ} \tag{3.1.5}$$

one can see that in the end the value of $T_{inp}$ ideally should converge to the values of $T_{targ}$. Therefore, from the both sides in Fig. 3.1.1 it is possible to use the values of $T_{targ}$ and train the NN. Of course, such simplification is only valid if one is building a SS surrogate (which is the case in this work and in [3]). At the same time, the iterative framework proposed in the paper [3] might be useful for unsteady surrogates with Recurrent Neural Networks (RNNs) [46]. Such opportunity might be explored in the future work.

### 3.2. Embedding a Neural Network in the Navier-Stokes Equations

As it is stated above, in this paper the SS solutions of the Navier-Stokes equations are considered. The architecture of the framework is presented in Fig. 3.2.1. The inputs to the NN are the field of velocities $u$ and $v$, while the aim is to build a surrogate model for the velocity-dependent dynamic viscosity, which is the output from the NN with some functional dependency $\mu = \mu(vel)$.

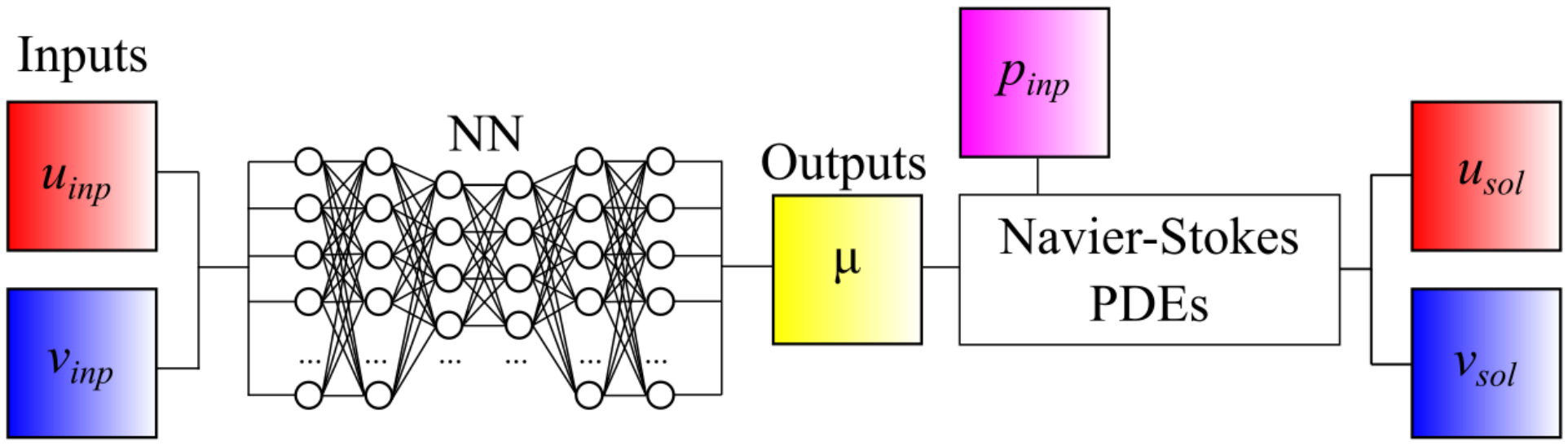

Fig. 3.2.1. Type 3 ML: a NN is embedded into the solution of the Navier-Stokes equations.

Again, we note that the values of the dynamic viscosity are not used as the training targets: instead, the training targets are the SS velocity fields $u_{targ}$, $v_{targ}$. Therefore, the cost function is defined as

$$C = \frac{1}{2}\frac{\sum\left(u_{sol_{i,j}} - u_{targ_{i,j}}\right)^2 + \sum\left(v_{sol_{i,j}} - v_{targ_{i,j}}\right)^2}{N_{data}} \tag{3.2.1}$$

Since the NN is embedded in the Navier-Stokes equations using the Tensorflow according to the Section 2.2, there is a "clear" traceable connection between input values $u_{inp}$, $v_{inp}$ and solutions $u_{sol}$, $v_{sol}$. Furthermore, since a SS flow is considered, the algorithm in Fig. 3.1.2 might be simplified according to the Fig. 3.2.2 and Eq. (3.1.5) (no global iterations since the values $u_{targ}$, $v_{targ}$ are used as the inputs). Such architecture reminds an encoder-decoder type of a NN. Note that the tensor $p_{inp}$ is also fed as an input field because it allows maintaining the connection between $u_{inp}$, $v_{inp}$ and $u_{sol}$, $v_{sol}$.

In the next section a case study is formulated to demonstrate the performance of the Type 3 ML with embedded DFNN in the Navier-Stokes equations.

| |
|---|
| **Prepare targets (which are also the inputs)**: velocity fields from the experiments (or simulations) $u_{targ}$, $v_{targ}$ **Prepare input**: pressure field $p_{inp}$ which corresponds to the steady-state solution $u_{targ}$, $v_{targ}$ |
| **For** *epoch* **in range** (0, $epoch_{max}$) **do** {iterations to train the NN} |
| Train the NN (by building a surrogate for $\mu(u_{targ}, v_{targ})$) to minimize $C$ (Eq. 3.2.2) |
| Use the trained NN to predict μ or to solve new datasets for the unknown $u$, $v$, $p$ |

Fig. 3.2.2. Simplified algorithm for a SS Type 3 ML.

## 4. Case Study No. 1: A Surrogate Model for Velocity-dependent Dynamic Viscosity

### 4.1 Mathematical Model

As the case study the 2D lid-driven cavity is chosen [47]. This is a famous benchmark problem: the dimensions of the cavity are 1×1 m; all 4 boundaries are motionless walls, except the upper one, which is moving with constant $x$-velocity $u_w = 1$ m/s. As a result, initially motionless fluid with density $\rho = 1$ kg/m$^3$ is accelerated; by choosing different Re numbers (or dynamic viscosities μ) the complex flows are modelled inside the cavity (e.g. Fig. 4.1.1 (*a*) demonstrates 3 vortices inside the cavity for Re = 100 and $\mu = \text{const} = 10^{-2}$ Pa·s). Fig. 4.1.1 (*b*)-(*d*) shows the SS solutions for pressure and velocity fields.

Thus, the mathematical model consists of the Navier-Stokes equations (2.1.1) and (2.1.2) with no-slip impermeable boundary conditions:

$$
\begin{gathered}
u(t, x=0, y) = u(t, x=1, y) = u(t, x, y=0) = 0 \\
u(t, x, y=1) = u_w \\
v(t, x=0, y) = v(t, x=1, y) = v(t, x, y=0) = v(t, x, y=1) = 0
\end{gathered}
\tag{4.1.1}
$$

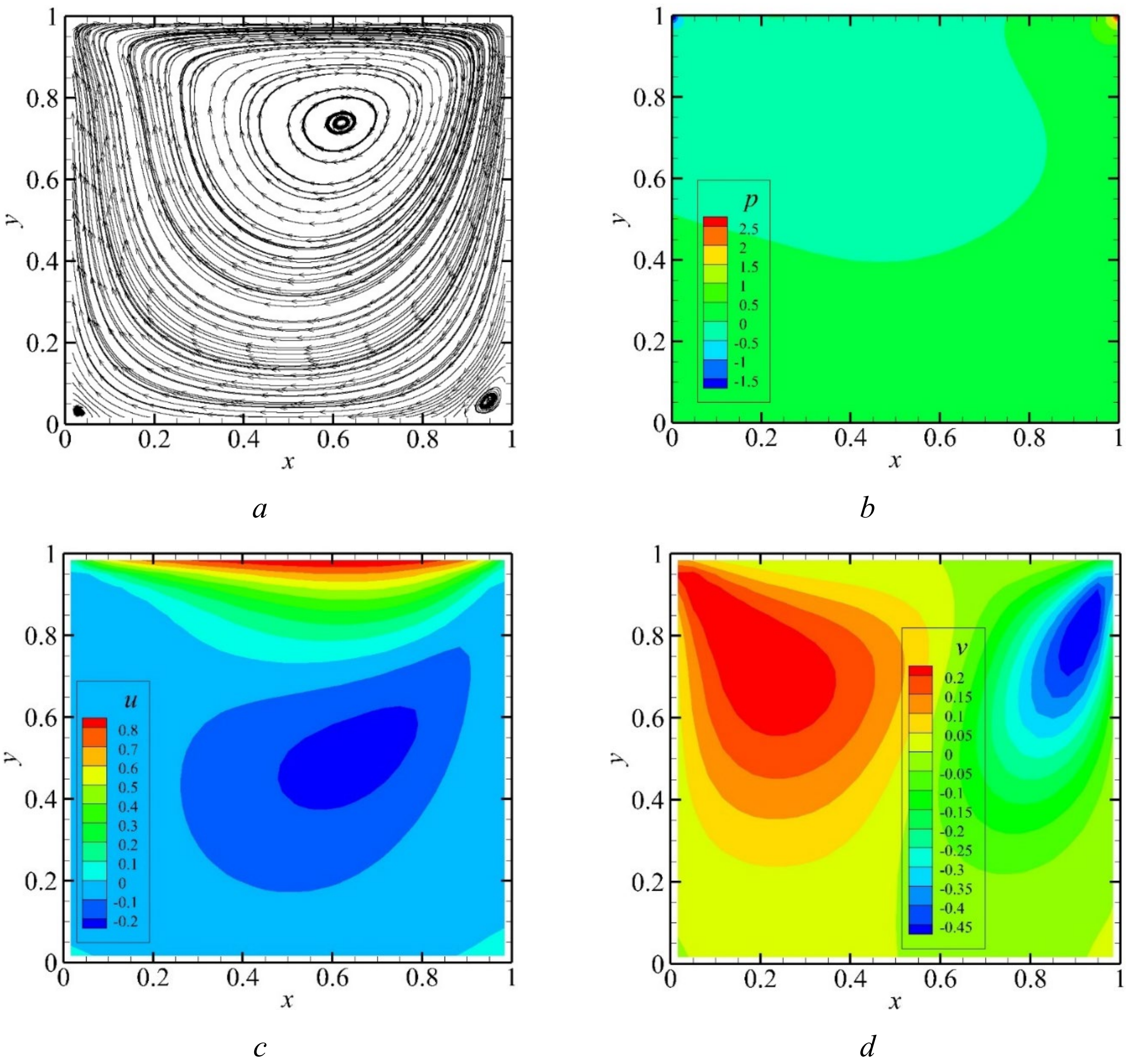

Fig. 4.1.1. SS solution of the 2D lid-driven cavity problem for Re = 100 ($\mu = 10^{-2}$ Pa·s); mesh 32×32 (including 2 ghost cells): (*a*) streamtraces (3 vortices are generated: in the central region and 2 bottom corners of the cavity), (*b*) pressure, (*c*) *u*-velocity, and (*d*) *v*-velocity.

### 4.2. Solution Verification

For numerical solution verification, velocity profiles along the vertical and horizontal centerlines of the cavity are compared with the results provided by Ghia U. et al. [47]; the results of comparison are shown in Fig. 4.2.1. The blue curves represent the solutions obtained using Python script with the Tensorflow library; the red curves are the Fortran solutions (developed for fast data generation, see Section 4.3); white dots are the results provided in [47]. As it could be seen, the results are very close; some minor deviations are caused by the control volume size (not enough mesh refinement for high Re numbers). Since later only small Re numbers are considered, the mesh 32×32 is adopted; such mesh allows to maintain quick solutions and reasonable accuracy.

The green curves represent the calculations with non-constant velocity-dependent dynamic viscosity according to the Eq. (4.2.1):

$$\mu = 4\mu_0 \exp\left(0.5vel\right) \tag{4.2.1}$$

where $\mu_0$ is the initial value of the dynamic viscosity (for a motionless fluid), *vel* is the velocity magnitude.

The adopted dependence $\mu(vel)$ noticeably changes the velocity profiles in the cavity and the aim is to "catch" this underlying dependence using the DFNN without having target values for $\mu$. This dependence for the dynamic viscosity might be considered as a simple model of a non-Newtonian fluid[3], which becomes more viscous at high velocities.

### 4.3. Training Data Generation

For fast training data generation analogous Fortran code is developed: Python is an interpretable language and it is very slow when it is necessary to perform iterations in loops, which are required to achieve the SS and solve the pressure equation (2.1.4). Additionally, since the Python with Tensorflow performs tensor operations, it is only possible to use the Jacobi iterations, while using the Fortran code it is possible to implement the Gauss-Seidel method, or successive overrelaxation (SOR) method, which substantially reduce the number of iterations for pressure-Poisson equation [48].

Comparison of the CPU time consumed by Fortran and Python is presented in Fig. 4.3.1. As it could be seen from the figure, the Fortran code is much superior than the Python script because the solution of the pressure equation requires hundreds of iterations on each time step and thousands of time steps to reach the SS.

[3]Strictly speaking, for a Non-Newtonian fluid the viscosity is non-constant and depends on stresses (not on velocity).

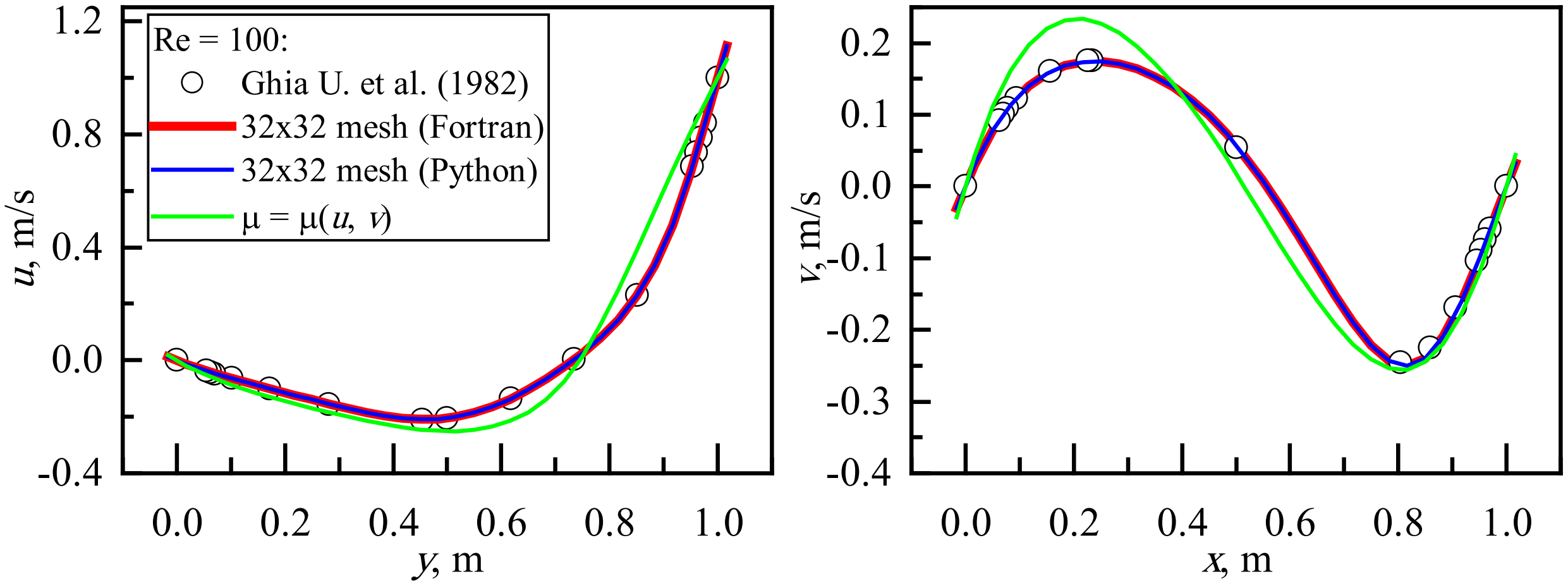

*a*

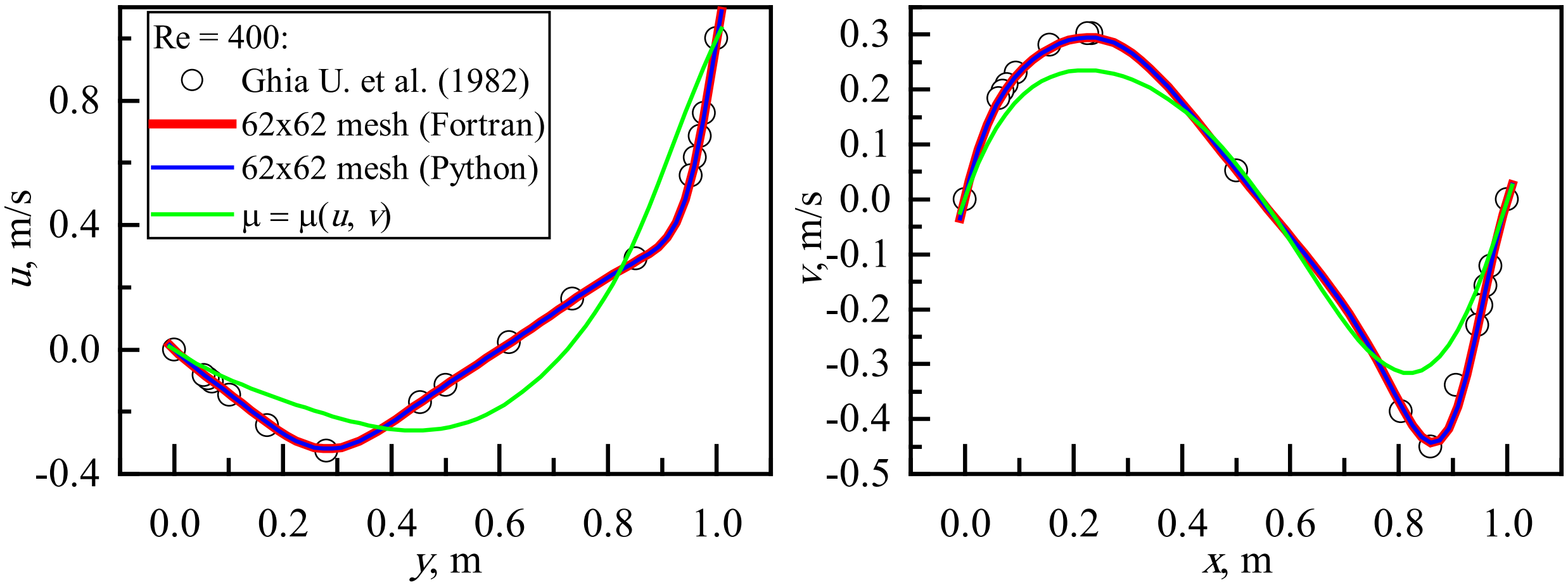

*b*

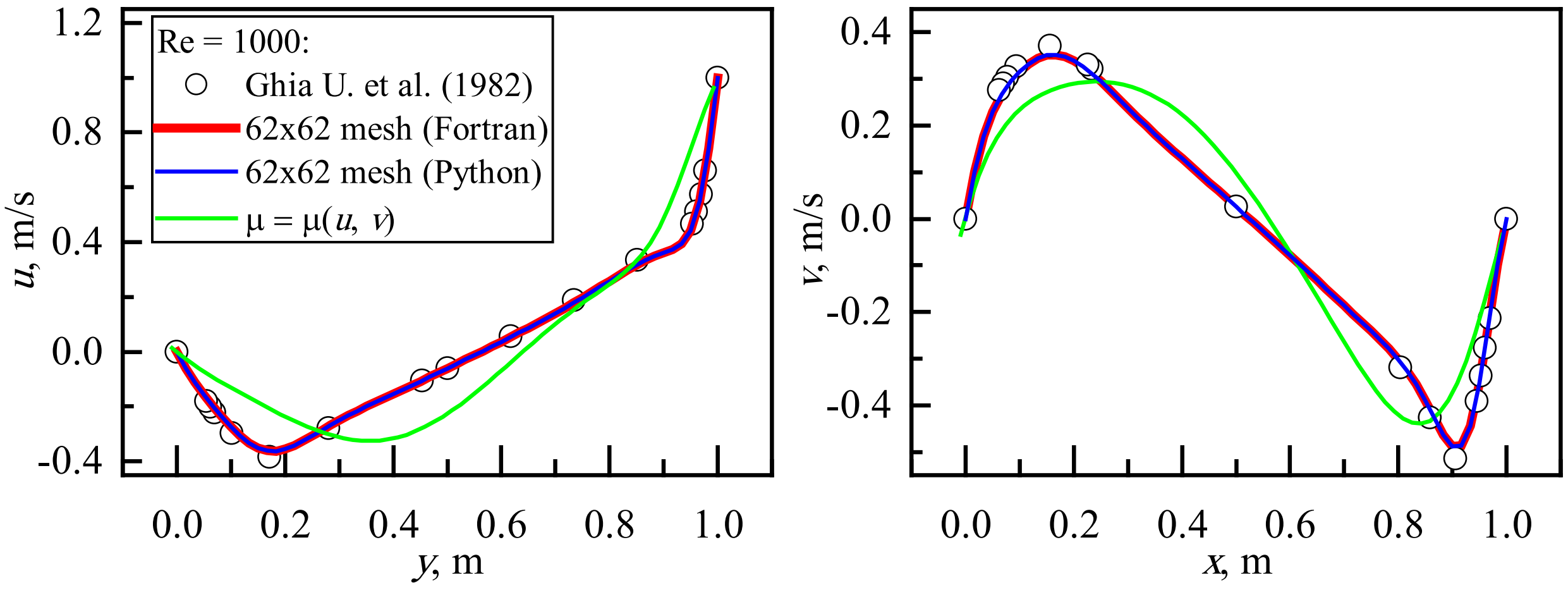

*c*

Fig. 4.2.1. Velocity profiles along the centerlines of the cavity: (*a*) Re = 100 ($\mu_0 = 10^{-2}$ Pa·s); (*b*) Re = 400 ($\mu_0 = 2.5\cdot10^{-4}$ Pa·s), and (*c*) Re = 1000 ($\mu_0 = 10^{-3}$ Pa·s).

At the same time, the Tensorflow session might be launched on a CPU or on a GPU[4]: the comparison shows that for a small mesh (less than ~60×60 nodes) the CPU solves the equations faster; however, for relatively large mesh the GPU outperforms the CPU in speed due to faster tensor operations). At the same time, it looks like the Tensorflow still does not provide a good platform for efficient parallelization in scientific simulations: element by element Fortran calculations require much less CPU time. Hopefully, the future development of the Python and Tensorflow[5] will allow employing GPU for faster (and easier) parallel solution of PDEs with large mesh sizes.

Note that data format used was float64 (double precision), which is not recommended for the Tensorflow (since it gives the best performance for float32 (single precision)), but float32 format does not have enough digits to solve the PDEs iteratively and reach a convergence criterion specified in Section 2.1.

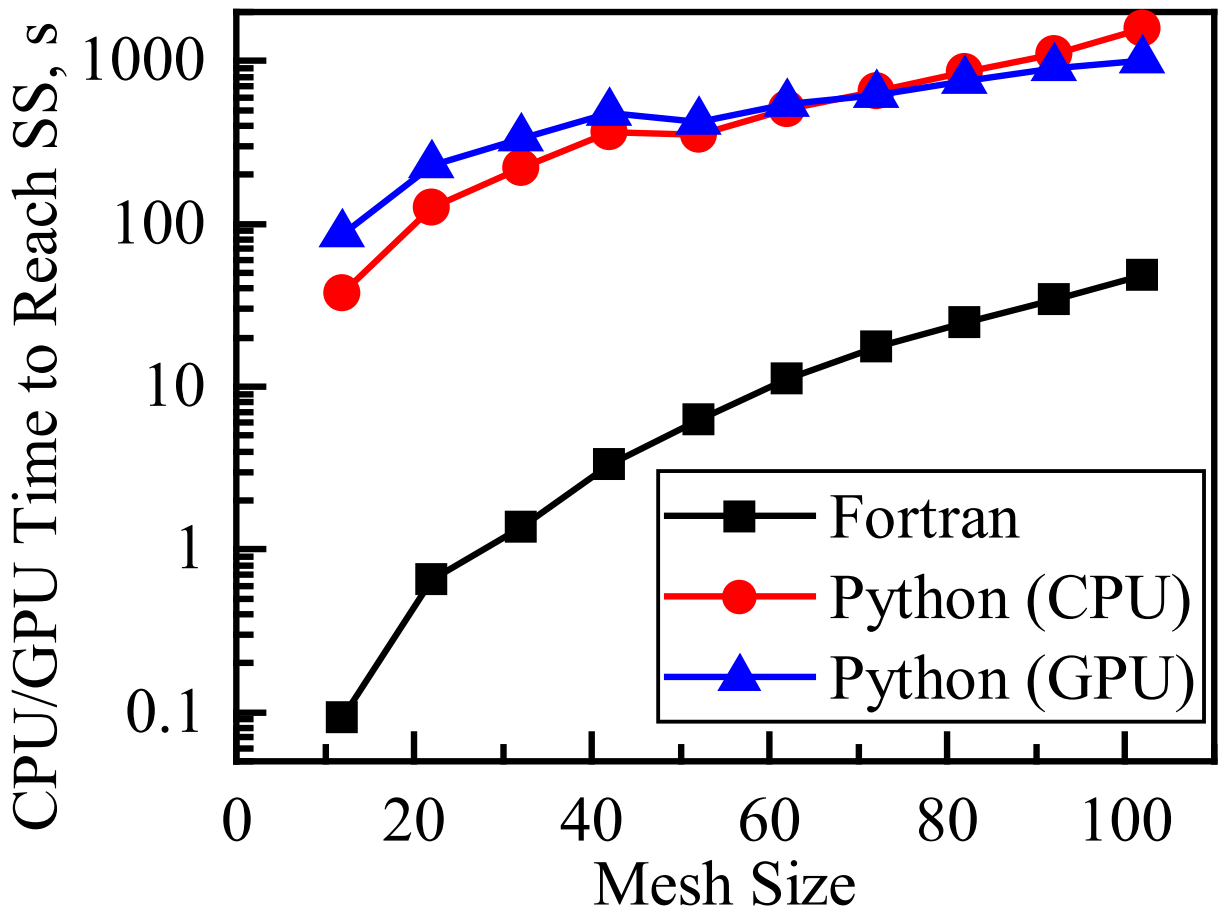

Fig. 4.3.1. Comparison of CPU/GPU time required to achieve SS by Python and Fortran programming languages.

Usually, when the lid-driven cavity is considered, the Re number (or viscosity, which is equivalent) is changed. However, since the aim is to build a surrogate for the viscosity, it is necessary to keep $\mu_0$ in Eq. (4.2.1) constant: instead of changing the viscosity, the upper wall velocity is varied in the range 0.50-1.50 m/s with a step 0.05 m/s (overall $N_{data} = 21$ training datasets are generated according to the Table 4.3.1).

---

[4]Calculations are performed using a usual laptop: Intel Core i7-6500 CPU 2.5 GHz and Nvidia GeForce GTX 950M GPU.

[5] In this work API version 1.15 is used, while the newest one is 2.3 [39].

Table 4.3.1. Training data.

| Dataset No. | Upper wall velocity $u_w$, m/s | $Re = \rho u_w L_x / \mu_0$ | $\mu_0$, Pa·s | ρ, kg/m$^3$ | $L_x = L_y$, m |
|---|---|---|---|---|---|
| 1 | 0.50 | 50 | | | |
| 2 | 0.55 | 55 | | | |
| … | …` | … | | | |
| 11 | 1.00 | 100 | $10^{-2}$ | 1.0 | 1.0 |
| … | … | … | | | |
| 21 | 1.50 | 150 | | | |

Fig. 4.3.2 shows the field of the dynamic viscosity for the training dataset No. 11 (Re = 100). It has maximum values in the upper region of the cavity where the velocity is higher due to the moving wall.

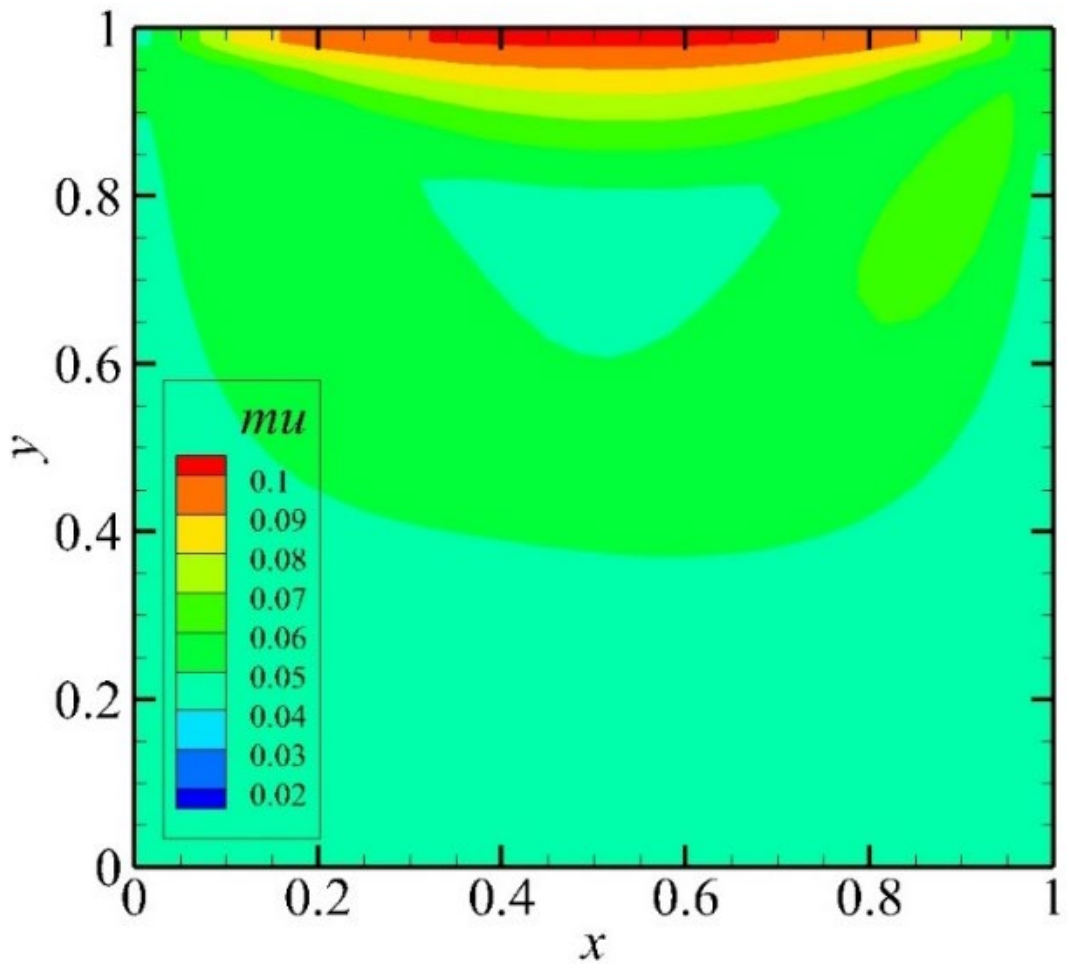

Fig. 4.3.2. Dynamic viscosity field for the training dataset No. 11.

## 4.4. Hyperparameters and Architecture of a Neural Network

A DFNN is used in this study, which has the input layer with 2×30×30 = 1800 neurons (fed with $u$ and $v$ fields), 4 hidden layers with 128 neurons in each layer, and the output layer with 30×30 = 900 neurons, which gives the dynamic viscosity field. Note that while the pressure and dynamic viscosity fields have dimensions 32×32, the ghost cells (which are used to set boundary conditions) are not used in the DFNN.

The sigmoid activation function σ(*z*) is used in all layers of the DFNN:

$$y_{out}(x_{inp}) = \sigma\left(\ldots\sigma\left(\sigma\left(x_{inp}\cdot W_1 + b_1\right)\cdot W_2 + b_2\right)\ldots\right)\cdot W_L + b_L \tag{4.4.1}$$

$$\sigma(x_{inp}) = \frac{1}{1+e^{-x_{inp}}} \tag{4.4.2}$$

where $y_{out}$ is the output from the NN, $x_{\mathrm{inp}}$ is the input, $W_l$ are the weights in the *l*th layer, $b_l$ is the biases of the *l*th layer, *L* is the number of layers in the NN.

The weights are initialized with the Xavier initialization [49], while the biases are generated using the normal distribution with the mean 0.0 and standard deviation 1.0. They are trained using the Adam algorithm [50] with a learning rate η = 0.01. To prevent the overfitting the dropout technique was tried with different rates [51], but it has not improved the performance of the DFNN.

Additionally, a CNN was trained, but it has shown worse performance than the DFNN probably due to the absence of the spatial structure in the data which might be extracted by the convolutional kernels. Perhaps, a CNN would be useful if dynamic viscosity dependence on stresses (not on absolute velocity) is considered.

## 5. Results and Discussion

The DFNN is trained inside the Navier-Stokes equations during 15000 epochs, which took approximately 10 minutes. It is interesting to note, that the originally proposed framework in [3] (Fig. 3.1.2) required much more time (hours and days) to train a NN for a much simpler heat conduction problem. It is also worth noting that since the cost function is connected with the output from the NN via the discretization scheme, the time step size (or CFL number) influences the training process: large CFL numbers increase the training speed, while the low CFL numbers decrease the training speed by reducing the gradient of the cost function (this effect might be "compensated" by a larger learning rate values).

Cost function dependence on the epoch number is presented in Fig. 5.1 (*a*). Additionally, to monitor the accuracy of the dynamic viscosity prediction by the DFNN, Root Mean Square Error (RMSE) is monitored (Fig. 5.1. (*b*)):

$$RMSE(\mu)=\sqrt{\frac{\sum\left(\mu_{nn}-\mu_{func}\right)^2}{(m-2)(n-2)}} \tag{5.1}$$

where $\mu_{func}$ is the functional dependence (4.2.1), $\mu_{nn}$ is the direct output from a NN.

After the training, the NN-based dynamic viscosity is used to predict the training data (velocities) and pressure field (Fig. 5.2): as it could be seen they are very close.

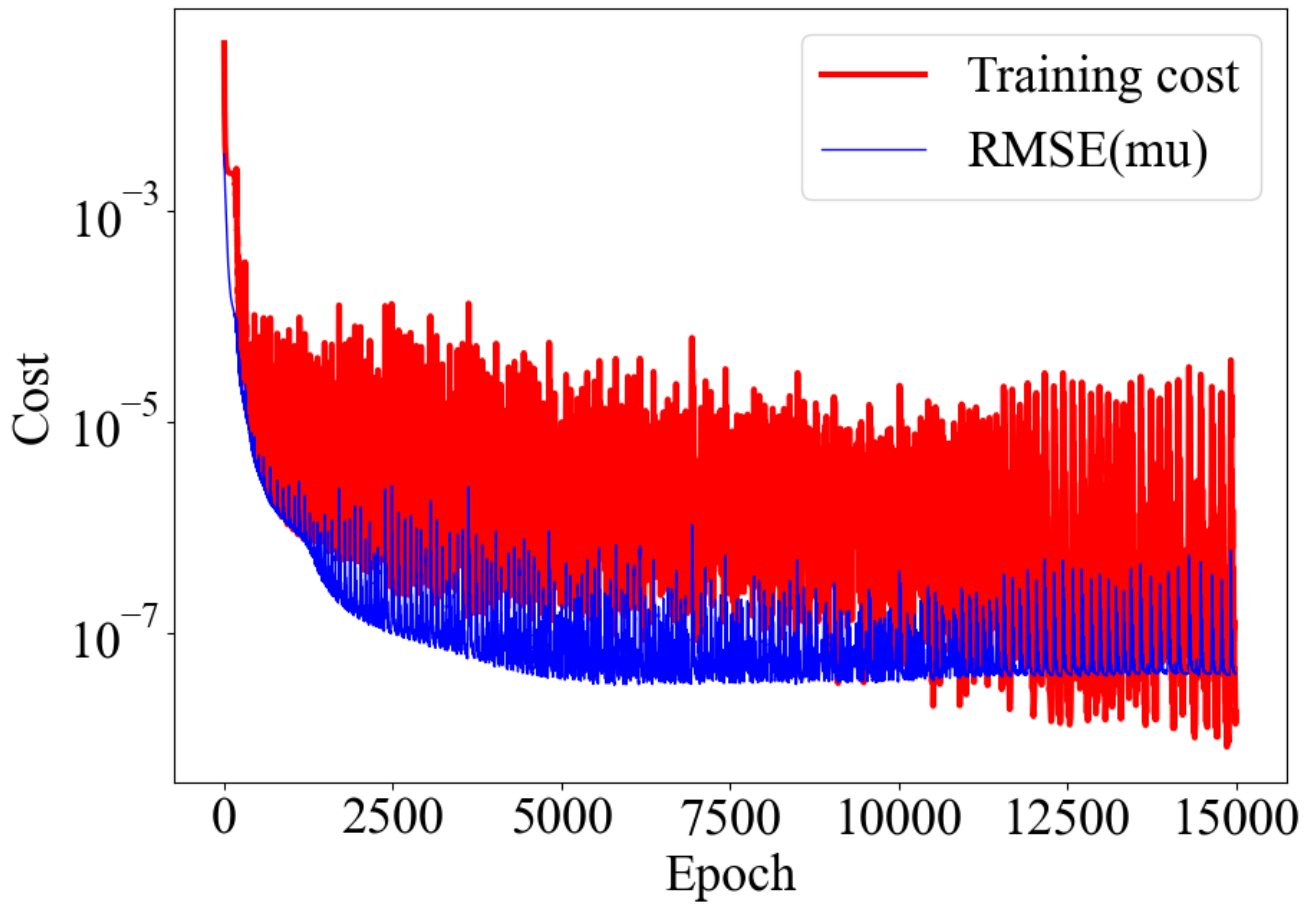

Fig. 5.1. Cost function and accuracy of the μ prediction versus epoch number.

Fig. 5.3 shows the NN-based dynamic viscosity, real dynamic viscosity fields (calculated using functional dependency Eq. (4.2.1)), as well as their difference for the dataset No. 11. The difference does not exceed 15% and has the highest value in the bottom corners of the cavity, where the vortices exist (see Fig. 4.1.1 (*a*)).

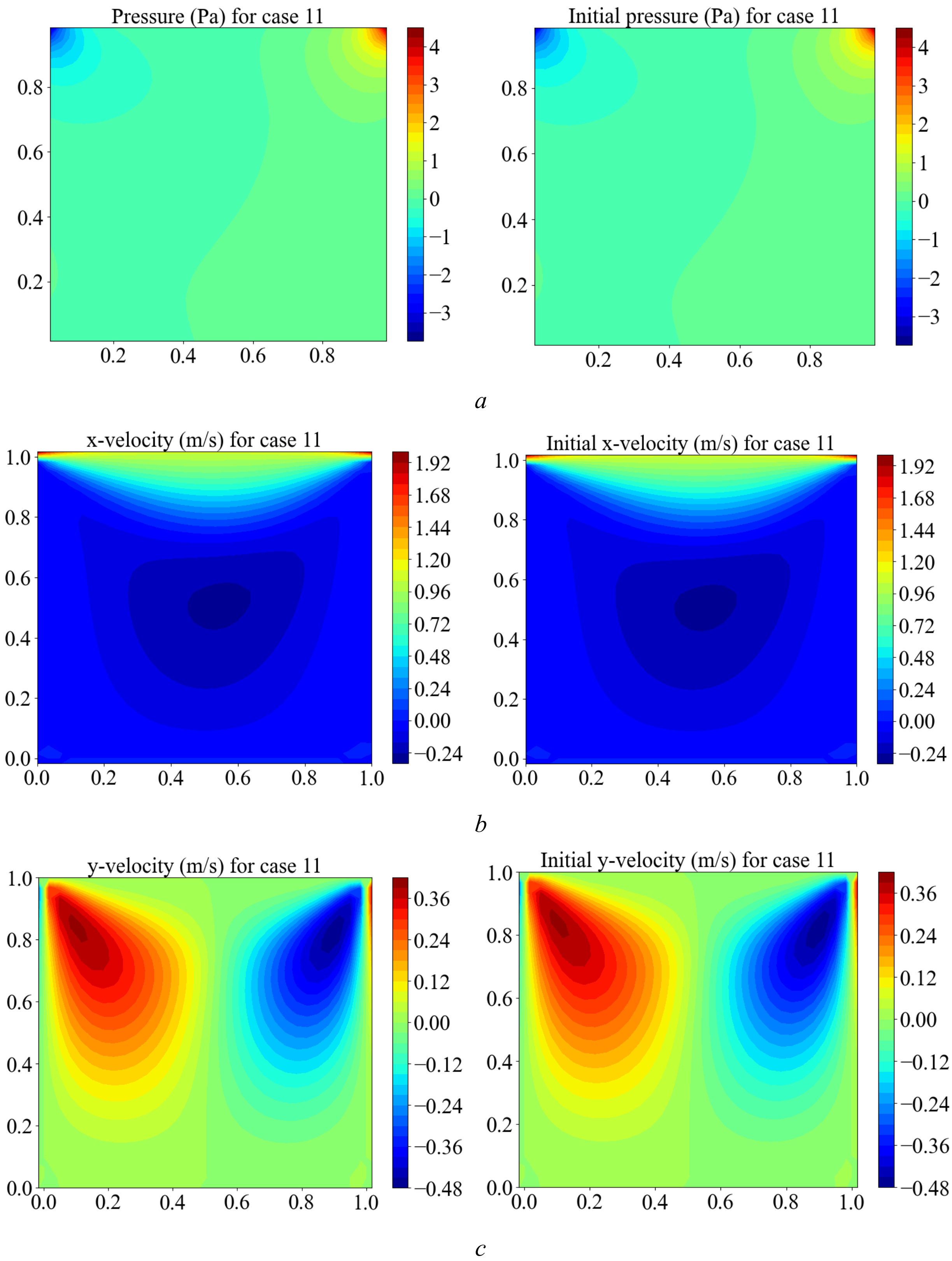

Fig. 5.2. Comparison of the solutions obtained with the NN-based dynamic viscosity (left) and target data (right): pressure (*a*), *u*-velocity (*b*), and *v*-velocity (*c*) for the dataset No. 11.

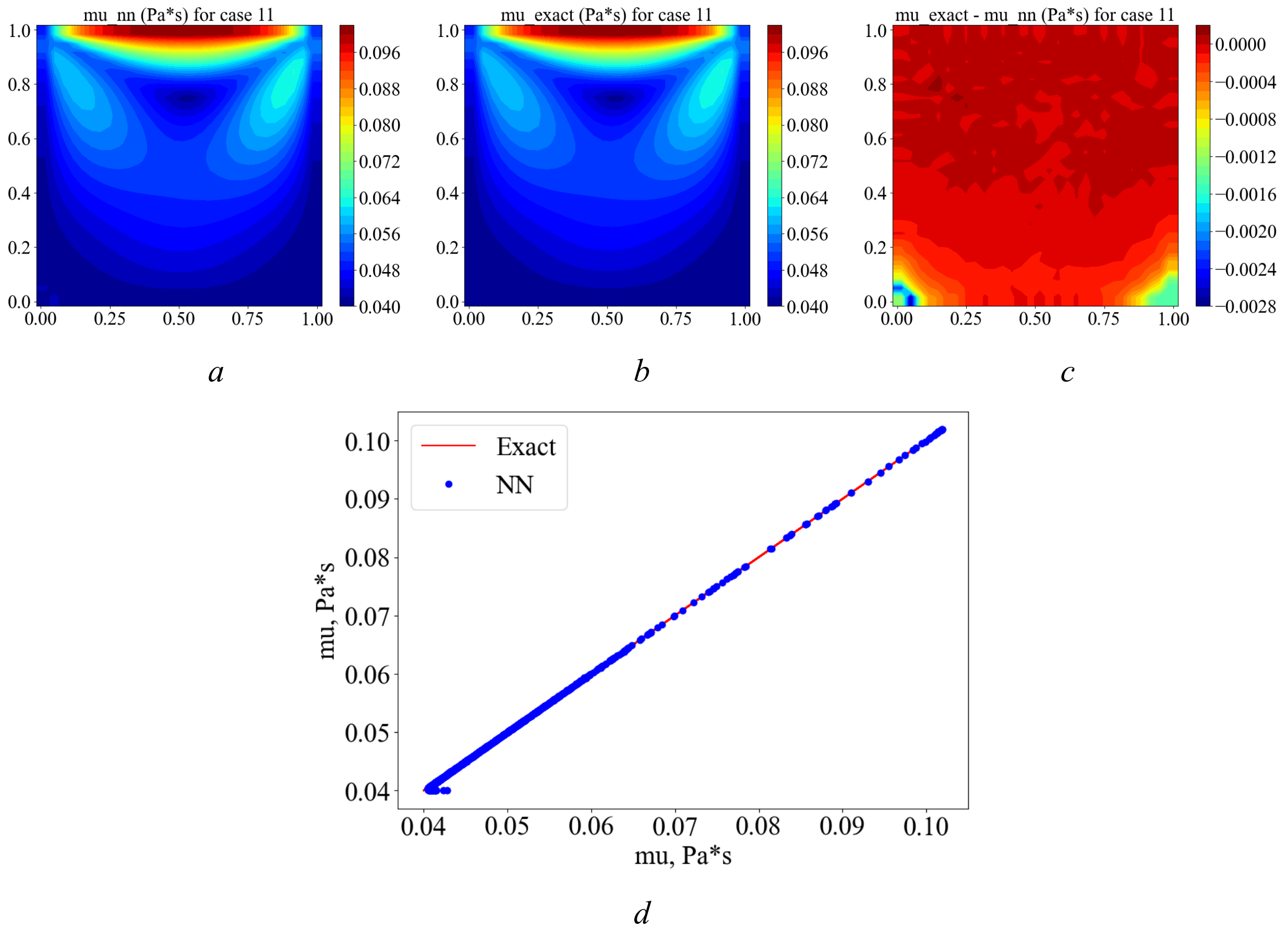

Fig. 5.3. NN-based dynamic viscosity (left), real dynamic viscosity (center), and the discrepancy between them (*c*) and (*d*).

However, the true test of a model is ability to make predictions in new situations. To check the performance of the DFNN, 3 different validation datasets are considered (2 extrapolation ones (the wall velocities are outside of the training range) and 1 interpolation dataset), according to Table 5.1.

Table 5.1. Validation data.

| Dataset No. | $u_w$, m/s | Re | Data coverage condition |
|---|---|---|---|
| 1 | 0.450 | 45 | Extrapolation |
| 2 | 1.025 | 102.5 | Interpolation |
| 3 | 1.550 | 155 | Extrapolation |

Fig. 5.4 shows the results of the profiles calculations for the validation dataset No. 2; the results are very close to each other (as well as for datasets No. 1 and No. 2, not shown). Fig. 5.5 demonstrates the deviations of the dynamic viscosity, velocities and pressure for all 3 validation datasets.

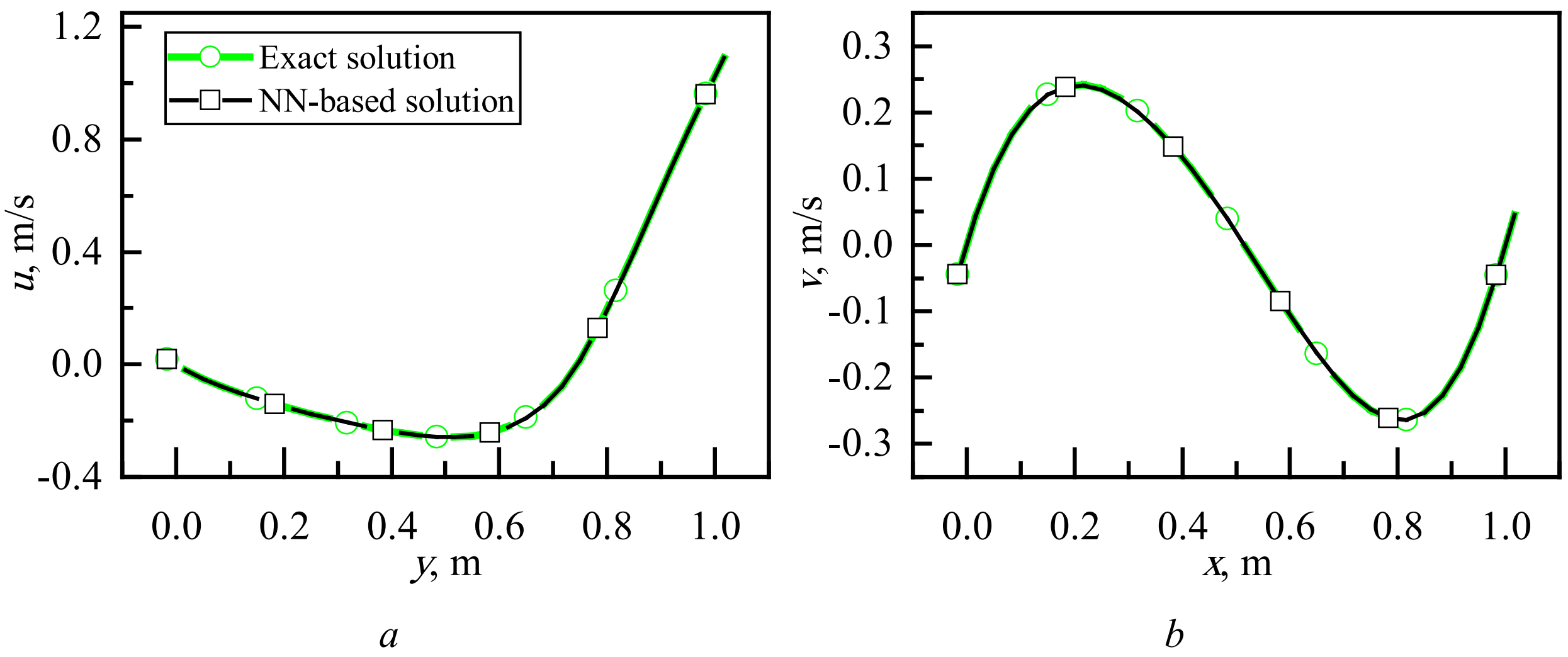

Fig. 5.4. Comparison of the exact velocity profiles with profiles calculated using the NN-based dynamic viscosity for validations dataset No. 2.

Appendix A demonstrates an implementation of Type 1 ML framework for a considered case study, when target data are available dynamic viscosity values. As it could be seen from Fig. 5.5 and Fig. A.2, Type 1 shows a slightly better performance and its computational complexity is much lower.

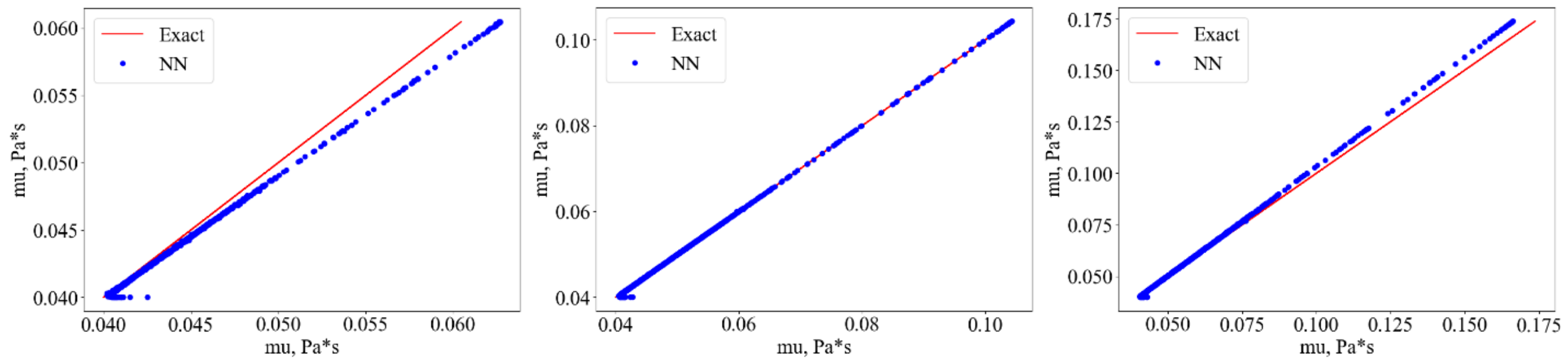

*a* *b* *c*

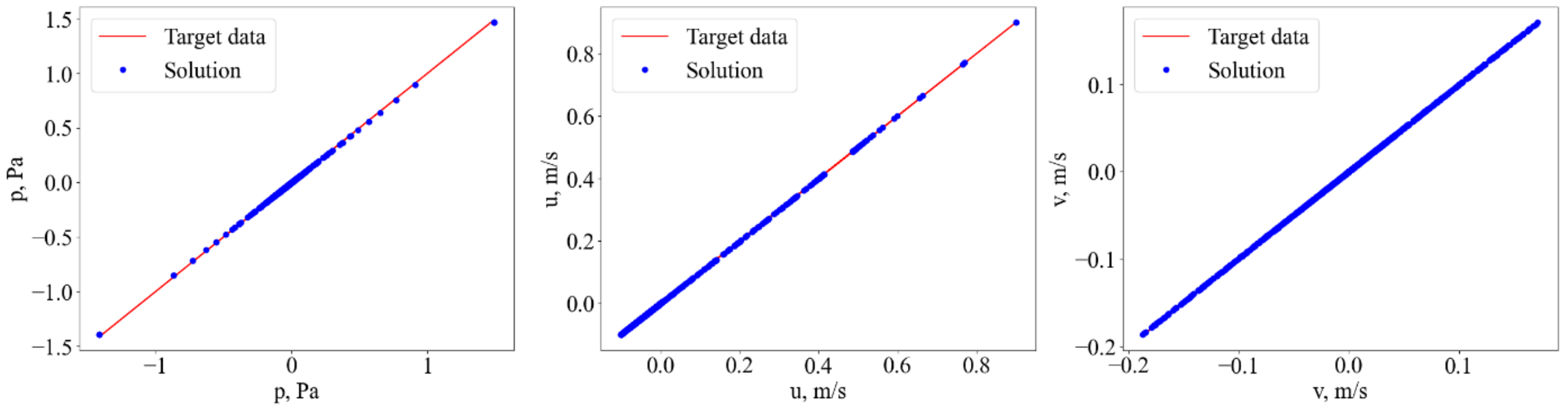

*a*

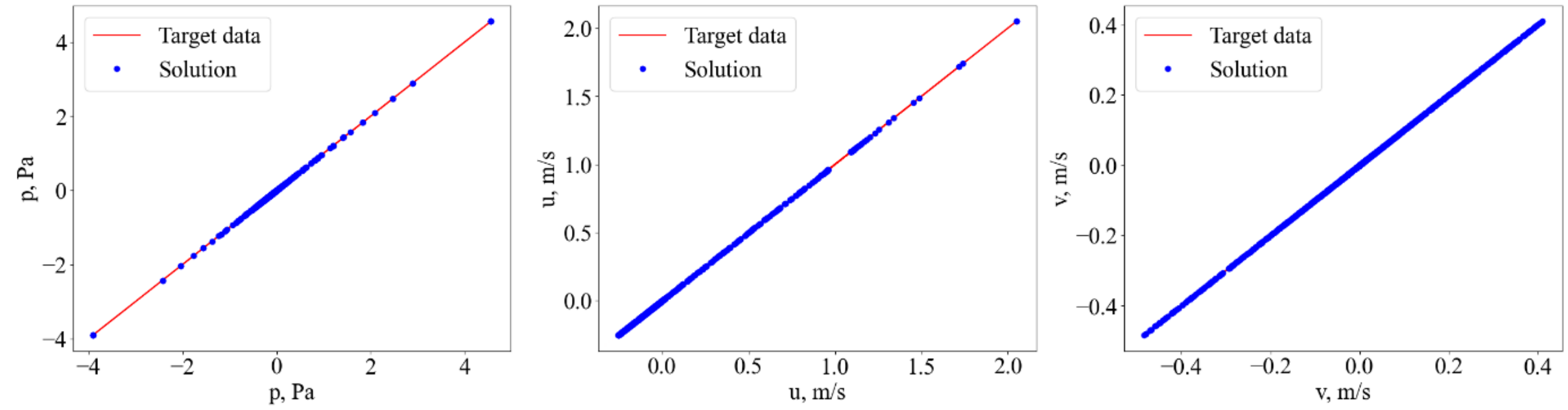

*b*

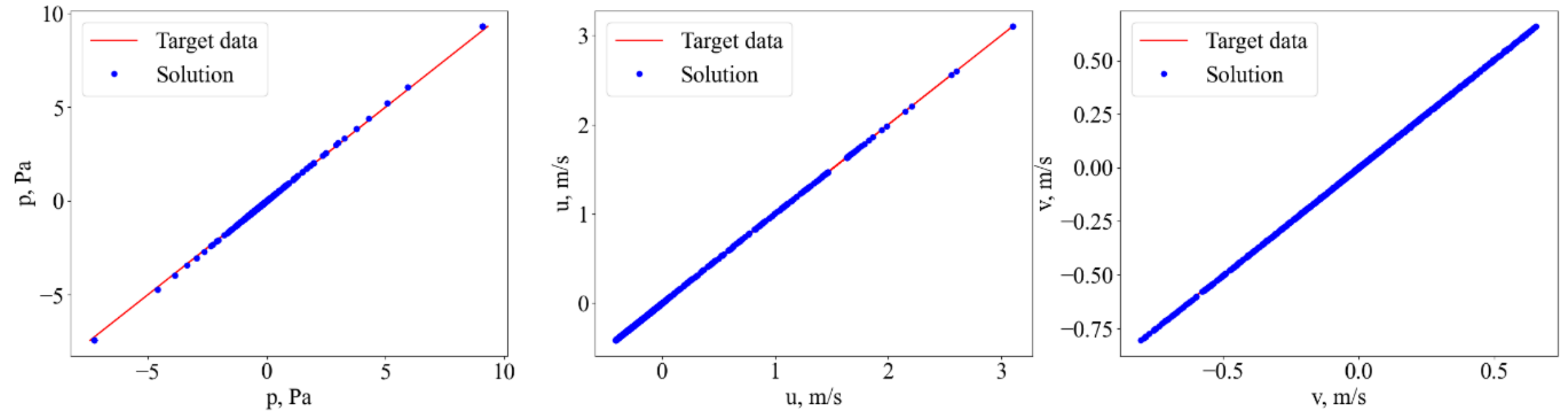

*c*

Fig. 5.5. Deviations of the dynamic viscosity, pressure, and velocities for validation dataset No. 1 (*a*), No. 2 (*b*), and No. 3 (*c*).

However, Type 3 suggests several advantages such as: (i) the target outputs for a NN might be unknown and can be recovered using knowledge-base (PDE integrated with a NN); (ii) it is not necessary to extract information from big data (labeled targets), instead it can be recovered by PDEs: for example, when building a CR for a turbulence model it is often necessary to extract physically interpretable data from the calculations such as Reynolds stress tensor and/or its characteristics, turbulent viscosity, etc., which is very time-consuming and may lead to the loss of relevant information; on the contrary, Type 3 may allow to use only QoIs (e.g. velocities) to recover necessary parameters (e.g. turbulent viscosity); (iii) there is no need to employ a physics- or scale-separation assumptions when building a CR; for instance, in multiphase flows modelling, many different forces on bubbles (or on droplets) are introduced, such as interfacial drag, lift force, virtual mass force, turbulent dispersion force, wall force [52]; it is clear that these forces are the models (e.g. there are still some argument on the physical meaning of the wall-force, which prevents bubbles from their accumulation near the walls). Indeed, nature does not separate this forces and Type 3 may allow to use a single force that may be represented by a NN trained in this manner on the DNS data.

The case study examined in this paper explicitly addresses demonstrates only point (i), while the other two (ii) and (iii) are the subjects for future work.

## 6. CONCLUSIONS

In this paper a framework to embed a NN in a system of PDEs is investigated. It is based on Type 3 (physics-integrated) ML framework proposed in [3], which is simplified (to significantly reduce the computational cost) and applied to the Navier-Stokes equations. The performance of the framework is demonstrated on a 2D lid-driven cavity problem with non-constant velocity-dependent dynamic viscosity: a DFNN was successfully trained to predict the dynamic viscosity without direct target data (values of viscosity). Instead, the viscosity field was recovered from the Navier-Stokes equations, which were solved using Chorin's projection method and the Tensorflow ML library. The proposed methodology for solution of the Navier-Stokes equations using the Tensorflow ML library is provided and may be adopted for further developments in the area of integration of ML with PDEs.

The developed framework is promising because (i) it allows to recover unknown physical values from the field variables if the governing equations for physics are known; (ii) it eliminates the necessity to extract physically-interpretable data from big data to train a NN; (iii) it eliminates the need to postulate a scale and physics separation. The future work will be aimed at further investigation of properties and opportunities made available by Type 3 ML framework.

### ACKNOWLEDGEMENTS

This research was performed with support of North Carolina State University Provost Doctoral Fellowship to the first author. The authors are also grateful to Dr. Chih-Wei Chang for his guidance in the implementation of the Type 3 ML framework.

## APPENDIX A. TYPE 1 IMPLEMENTATION FOR THE CASE STUDY NO. 1

Type 1 is a classical framework for development of DD CRs. It is applicable when target data for training are available and PDEs and CRs are scale-separable. In this case the cost function directly connects outputs from a NN and labeled target data (compare with (3.2.2)):

$$\text{Cost} = \text{RMSE}(\mu) = \sqrt{\frac{\sum\left(\mu_{nn} - \mu_{targ}\right)^2}{(m-2)(n-2)}} \tag{A.1}$$

Here the DFNN is trained with the same parameters as in Type 3 demonstration to ensure a correct comparison (see Section 4.4). Fig. A.1 shows the cost on training data (Table 4.3.1) and on validation (evaluation) data (Table 5.1). Fig. A.2 shows the deviations of the dynamic viscosity.

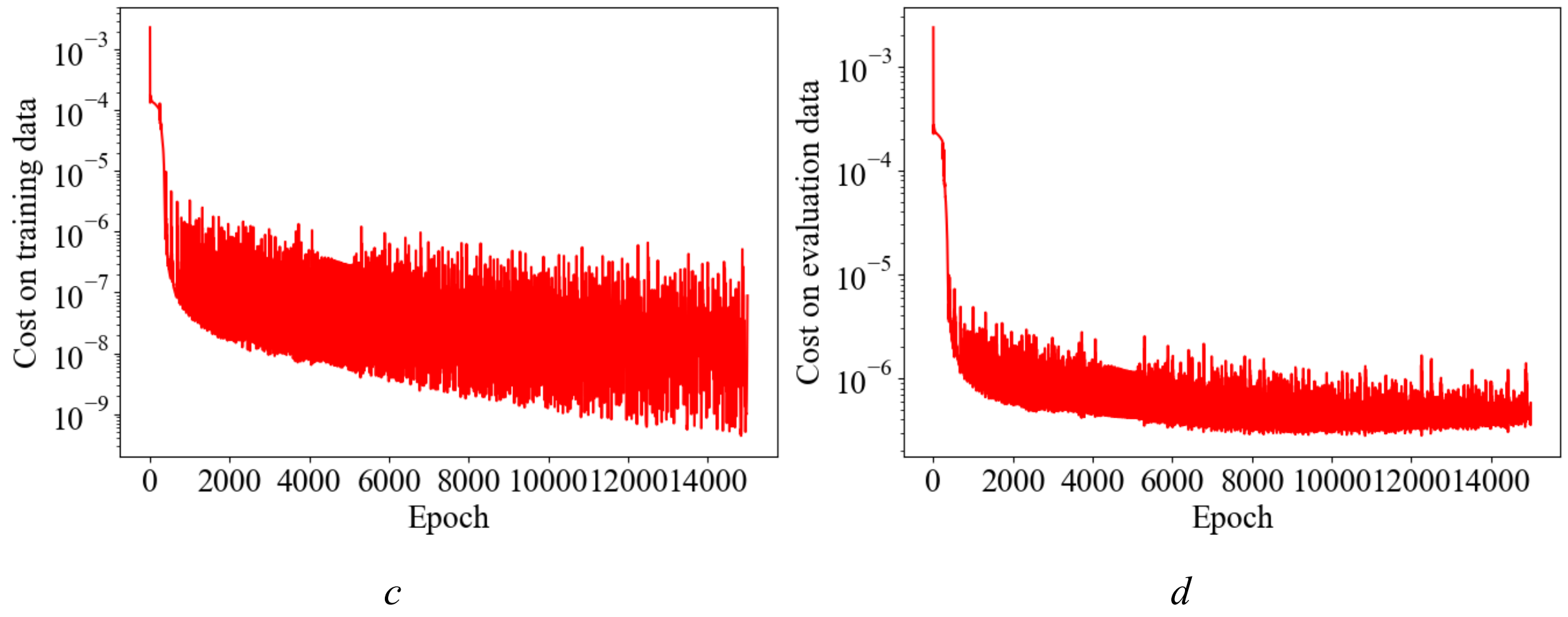

Fig. A.1. Cost function on the training data (Table 4.3.1) and on the validation data (Table 5.1).

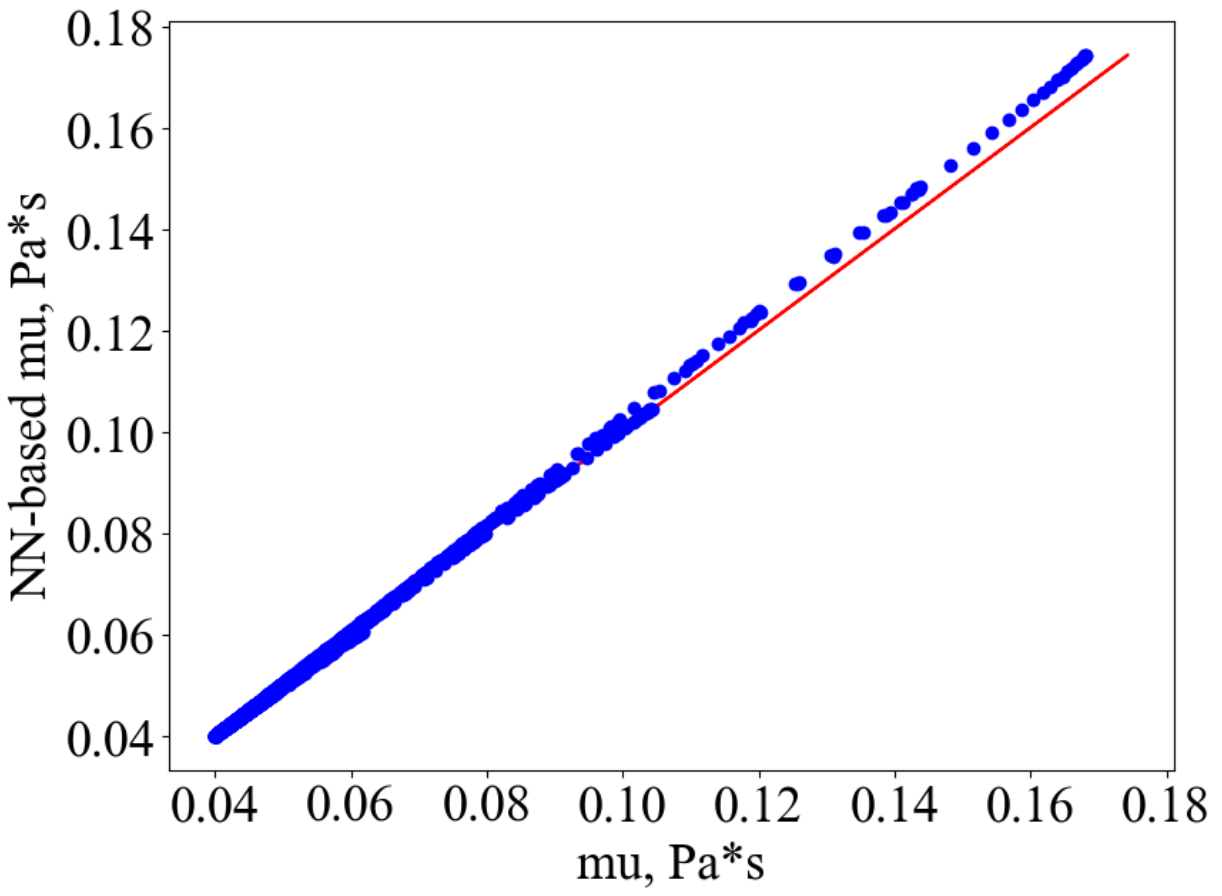

Fig. A.2. Deviations of dynamic viscosity for Type 1 ML for validation datasets.